\documentclass[12pt,preprint]{aastex}

\shorttitle{Wide-field weak lensing by RXJ1347--1145}
\shortauthors{Kling et al.}

\begin{document}

\title{Wide-field weak lensing by RXJ1347--1145}

\author{Thomas P. Kling}
\affil{Dept. of Physics, Bridgewater State College, Bridgewater,
MA 02325}

\author{Ian Dell'Antonio}
\affil{Dept. of Physics, Brown University, Providence, RI 02912}

\author{David Wittman\altaffilmark{1} and J.~Anthony
Tyson\altaffilmark{1}} \affil{Bell-Laboratories Lucent
Technologies, Murray Hill, NJ 07994}

\altaffiltext{1}{Visiting astronomer, Cerro Tololo Intra-American
Observatory.  The Cerro Tololo Intra-American Observatory is
operated by the Association of Universities for Research in
Astronomy under a cooperative agreement with the National Science
Foundation.}

\begin{abstract}

\noindent We present an analysis of weak lensing observations for
RXJ1347--1145 over a $43\,' \, \times \, 43 \, '$ field taken in
$B$ and $R$ filters on the Blanco 4m telescope at CTIO.
RXJ1347--1145 is a massive cluster at redshift $z=0.45$.  Using a
population of galaxies with $20<R<26$, we detect a weak lensing
signal at the $p<0.0005$ level, finding best-fit parameters of
$\sigma_v=1400^{+130}_{-140}$ km s$^{-1}$ for a singular
isothermal sphere model and $r_{200} = 3.5^{+0.8}_{-0.2}$ Mpc with
$c = 15^{+64}_{-10}$ for a NFW model in an $\Omega_m = 0.3$,
$\Omega_\Lambda = 0.7$ cosmology. In addition, a mass to light
ratio M/$L_R$ $=90 \pm 20~M_\odot / L_{R\odot}$ was determined.
These values are consistent with the previous weak lensing study
of RXJ1347--1145 by \citet{FT}, giving strong evidence that
systemic bias was not introduced by the relatively small field of
view in that study. Our best-fit parameter values are also
consistent with recent X-ray studies by \citet{allen2002} and
\citet{ettori} but are not consistent with recent optical velocity
dispersion measurements by \citet{cohen-kneib}.

\end{abstract}


\section{Introduction}

In this paper, we present recent weak lensing observations and
analysis of RXJ1347-1145, the most luminous galaxy cluster in the
ROSAT All Sky Survey \citep{schindler95}, which lies at a redshift
of $0.45$. RXJ1347-1145 has been the subject of numerous
observations, including a previous weak lensing study by
\citet{FT}, and several X-ray and optical studies. The primary
purpose of the wide-field weak lensing observation presented in
this paper is to test and confirm the earlier gravitational
lensing observations of \citet{FT}.

Early gravitational lensing studies of RXJ1347-1145 determined a
different total mass than early X-ray studies. Such observed
disparities for a large and luminous cluster like RXJ1347-1145
raised the possibility of potential systemic problems in
methodology.  As galaxy cluster mass distributions are a useful
probe for dark matter models, the lensing, X-ray, and optical
communities independently sought better observations of this
cluster.

\citet{FT} studied RXJ1347-1145 through weak lensing analysis of
data taken on the Blanco 4m telescope at CTIO using the prime
focus CCD camera in 1995. Using an approximately $14' \times 14'$
field of view, they found that the mass interior to a 1 Mpc radius
was $1.7\, \pm\, 0.4\, \times\, 10^{15}\, M_\odot$. Assuming an
isotropic velocity distribution, this mass corresponded to a
velocity dispersion of $1500\pm 160$ km s$^{-1}$.

The results of \citet{FT} were not consistent with the X-ray
results of \citet{schindler97}, who concluded that the temperature
of the cluster was $9.3^{+1.1}_{-1.0}$ keV. As their study
presented no evidence that there was a temperature drop from the
center of the cluster, \citet{schindler97} assumed a constant
temperature to find that the total mass within 1 Mpc was $5.8
\times 10^{14} M_\odot$.

There were several, independent concerns regarding these weak
lensing and X-ray studies that led to further observations.  The
primary concern for the X-ray studies was the constant temperature
assumption in the presence of an apparent large cooling flow.  For
the weak lensing studies, there was a possibility that the
relatively small field of view would introduce a systematic bias,
as the outer weak lensing measurements directly influence the
predicted interior mass.

Since 1997, several X-ray studies of RXJ1347-1145 have been
carried out, most recently Chandra \citep{allen2002} and BeppoSAX
\citep{ettori} observations.  The BeppoSAX observations of
\citet{ettori} determined that the $3\sigma$ lower bound of the
X-ray temperature of RXJ1347-1145 was $12.1$ keV, ruling out the
$9.3^{+1.1}_{-1.0}$ keV temperature measurement of
\citet{schindler97} at the $99.99\,\%$ confidence level. The
Chandra observations of \citet{allen2002} determined the
temperature of the cluster to be $12.2\pm0.6$ keV; in rough
agreement with the predictions of \citet{ettori}.

This temperature measurement of \citet{allen2002} is consistent
with the analysis of ASCA data by \citet{allen1998}.  The
temperature is also confirmed by two measurements of gas
temperature using the SZ effect -- those of \citet{pointecouteau}
and \citet{komatsu}.  The mass modelling consistent with
\citet{allen2002} is also consistent with the weak lensing
observations of \citet{FT}.

The recent X-ray measurements of \citet{allen2002} are not,
however, in agreement with the recent Keck spectroscopic survey
conducted by \citet{cohen-kneib}. Surveying 47 spectroscopically
confirmed members of RXJ1347-1145, \citet{cohen-kneib} determine a
central mass that is significantly lower than the X-ray or
previous lensing studies. \citet{cohen-kneib} summarize, in some
detail, the current X-ray and strong lensing predictions and the
past weak lensing predictions in Table 4 of their paper.

The weak lensing observations presented in this paper are based on
images of an approximately $43' \times 43'$ usable field of view
taken in both the $R$  and $B_j$ bands on the Blanco 4m telescope
at CTIO in 2000 with with light-to-moderate cirrus and $0.8-1''$
seeing.  Although the imaging depth of this study is less than
that of \citet{FT}, the observations presented here include an
approximately nine times larger field of view with better seeing.

The primary goal of this wide-field gravitational lensing study is
to rule out any systemic bias introduced by the smaller fields of
view in \citet{FT}.  A secondary goal of this paper is to
investigate the discrepancy between the current X-ray and optical
velocity dispersion measurements of RXJ1347-1145.

Because the observations presented here were designed for weak
lensing analysis, they do have sufficient angular resolution to
definitively judge either the X-ray or optical studies, which
focus in detail on the central region of the cluster. However,
mass models from this wide-field weak lensing observation can be
compared with other recent studies and will add to the overall
understanding of RXJ1347-1145.


\section{Observations}

The observations were carried out on March 10, 2000 using the
Blanco $4$-meter telescope at the Cerro Tololo Inter-American
Observatory with the MOSAIC II $8$k~$\times~8$k camera.  A total
of ten $720$-second R and eight $420$-second $B_j$ exposures were
taken with light-to-moderate cirrus and $0.8-1''$ seeing.  The
usable field of view was approximately $43'$ by $43'$, given the
pixel size of $0.258''$ and a large dithering pattern needed for
good night sky flats.

The images were processed using the IRAF package {\tt{mscred}} up
to the flatfielding; the image registration and stacking was made
with a preliminary version of the Deep Lens Survey (DLS) pipeline
software \citep{wittman}, which convolves the images with a
circularizing kernel \citep{FT} during the combining step,
resulting in a circular point spread function (PSF). Object
catalogs were generated with {\tt{SExtractor}} \citep{bertin}, and
the shape parameters were re-measured using {\tt{ellipto}}
\citep{bernstein} for all the objects.

Because the night was not photometric, we were unable to obtain a
precise zeropoint. Furthermore, we had too few unsaturated USNO
stars to permit zeropointing by matching to the USNO-A catalog. We
were able to achieve a rough calibration by matching photometry
with that for objects from the previous weak-lensing observations
of \citet{FT}, which used the same telescope (but a different
camera). Because in this paper we are interested in the shape
information primarily, and the photometry and colors are used only
to construct {\emph{relative}} color-magnitude diagrams for object
selection, our results are not affected by this zeropoint
ambiguity. However, we caution against using the data presented
here for absolute photometry.


\section{PSF Anisotropy \label{PSF}}

In general, a spatially variable intrinsic point spread function
anisotropy (PSF) will be present in our image.  This anisotropy
must be removed before determining the weak-lensing induced shear.
We corrected for PSF anisotropy by finding fourth-order
polynomials of the spatial position that circularize the PSF under
convolution with the original image, as in \citet{FT}.

To determine the convolution kernel, we measured the initial
moments, $\{I_{xx}, I_{yy}, I_{xy} \}$, for the stars in our image
and solved for a kernel that, when convolved with the original
image, yielded star moments $I^k_{xx}=I^k_{yy}$, and $I^k_{xy}=0$.

\begin{figure}
\plotone{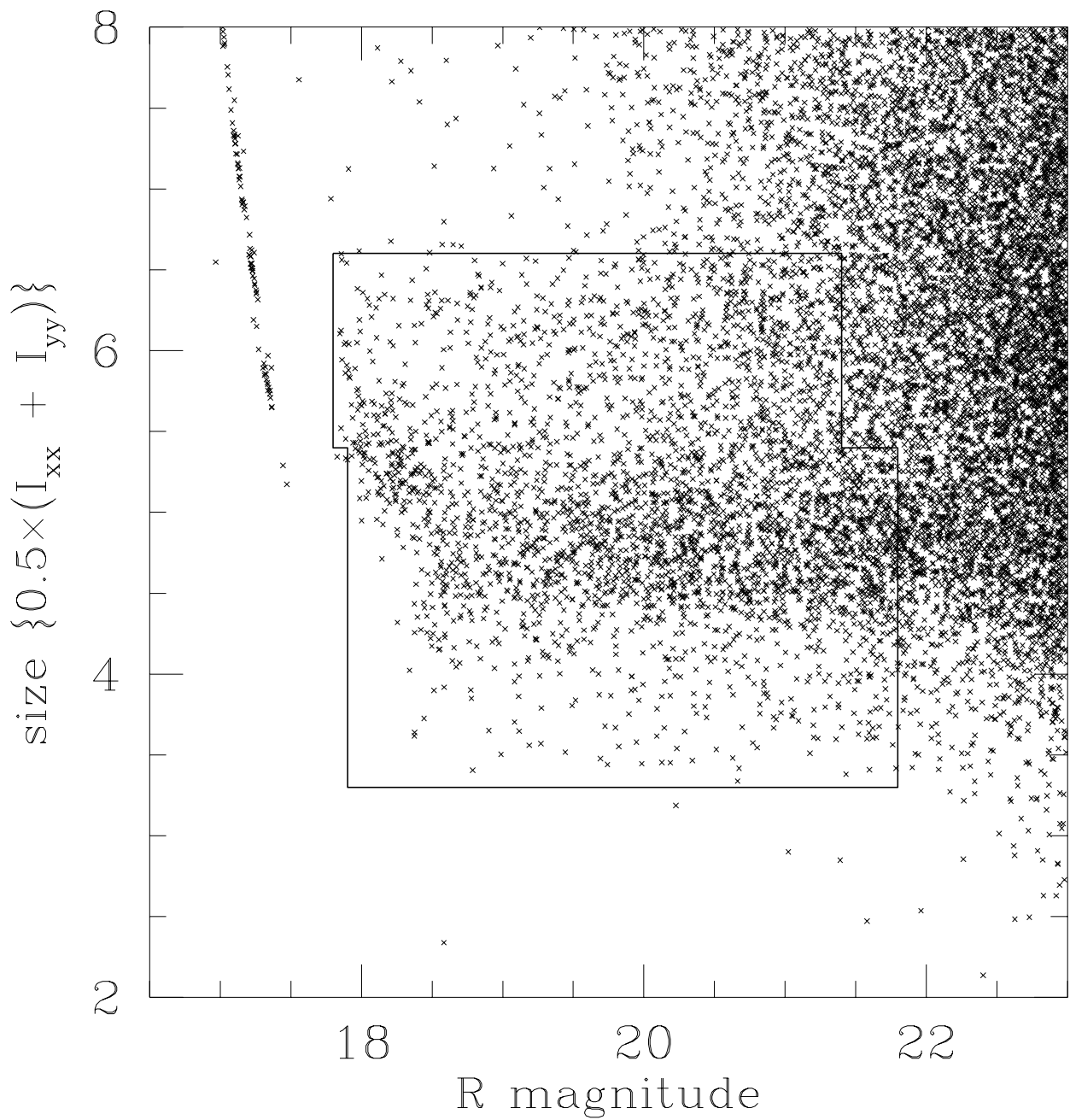} \caption{A plot of size versus $R$-magnitude
for well-defined objects in our sample. To determine a convolution
kernel that circularized the PSF, potential stars were selected
from the boxed region using a spatially variable criterion. Note
that many objects in the boxed region were not included because
they do not meet the additional spatial location criteria.
\label{star_select} }
\end{figure}

Potential stars were selected from objects in the box of
Fig.~\ref{star_select}.  Because the PSF size varied vertically
across the image, any single $R$ magnitude versus size criteria
either left regions of the image underpopulated or selected too
many galaxies.  Therefore, potential stars were selected using an
$R$ magnitude versus size criterion that varied across the image
to ensure that the star population contained members from all
regions.  Note that this necessarily implies that many of the
objects in the box of Fig.~\ref{star_select} were not included in
the initial star selection.

\begin{figure}
\plotone{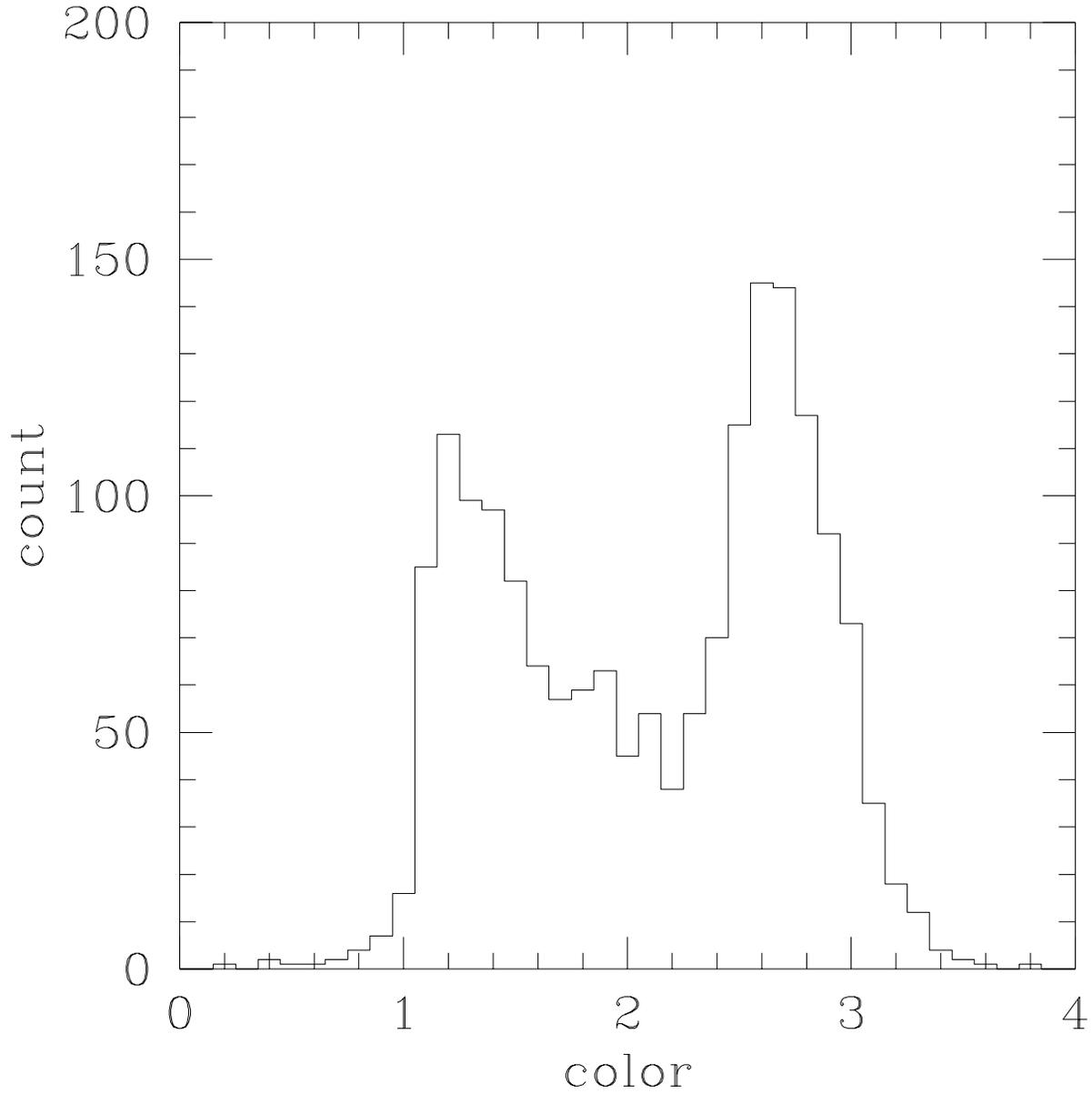} \caption{A histogram indicating the number of
objects of a given color in the potential star sample in the field
of view of RXJ1347-1145.  The bi-modal distribution indicates that
the potential star sample might be contaminated by blue galaxies.
A color cut of $B-R>2.0$ was applied to limit the sample to likely
stars. \label{starcolor} }  \end{figure}

As shown in Fig.~\ref{starcolor}, this potential star sample was
contaminated by a large number of unresolved, bluish galaxies.
Therefore, a color cut was made to exclude objects with
{\emph{observed}} color $B_j-R>2.0$. After this color selection,
we obtained a sample of 1063 non-saturated stars, or approximately
$0.57$ stars per square arc~minute.

The average (uncorrected) ellipticity,

\begin{equation}
\left<\bar{e}_1,\, \bar{e}_2\right> = \left< \frac{I_{xx} -
I_{yy}}{I_{xx}+I_{yy}},  \frac{2I_{xy}}{I_{xx}+I_{yy}} \right>,
\label{ellip_def}
\end{equation}

\noindent for these stars was

\begin{eqnarray}
\bar{e}_1 &=&-0.0030 \pm 0.0008, \nonumber \\
\bar{e}_2 &=&~0.0010 \pm 0.0005. \label{orig_e}
\end{eqnarray}

\noindent This sample had an average PSF size of $(1/2)(I_{xx} +
I_{yy}) = 0.631 \pm\, 0.003$ square arc seconds.

Applying our spatially dependent, fourth-order convolution kernel
to our original $R$ image was successful in circularizing the
stars in our sample. After the convolution, our 1063 star sample
is characterized by ellipticities

\begin{eqnarray}
\bar{e}_1 &=& 0.0007 \pm 0.0008, \\
\bar{e}_2 &=& 0.0000 \pm 0.0005. \label{final_e}
\end{eqnarray}

\noindent The star sample's PSF size increased slightly to $0.639
\pm 0.003$ square arc seconds, as is expected under a
circularizing convolution.


\section{Weak lensing analysis}

In this section, we discuss issues related to object selection,
systematic errors, and weak lensing detection.  We consider two
object selection criterion, and show that by maximizing the number
of objects available for analysis, a highly significant weak
lensing detection is made that is not influenced by systematic
errors.

\subsection{Object selection}

Selection of potential galaxies in the background of RXJ1347--1145
was achieved using a variety of cuts used to eliminate stars and
saturated, poorly resolved, or foreground objects.  Such objects
would reduce the strength of the weak lensing detection because
their shape is not correlated with the gravitational lensing of
the cluster.

First, the shapes of all objects detected by {\tt{SExtractor}}
were determined by the program {\tt{ellipto}}, which is discussed
in \citet{bernstein}.  {\tt{Ellipto}} carefully separates objects
and measures moments based on user defined criteria beginning with
the shape and centroid measurements from {\tt{SExtractor}}.
Because the seeing in the $B_j$ image was poorer than the $R$
image, the shapes measured by {\tt{ellipto}} were based on the $R$
image only.

Using a set of difficult to deblend objects from our image, we
determined that to achieve optimal shape measurements, the
centroid of a given image could not be allowed to wander more than
$30\%$ of the size of the object as {\tt{ellipto}} determined the
object's shape. Objects whose shape {\tt{ellipto}} could not
determine were excluded from further analysis.

\begin{figure}
\plotone{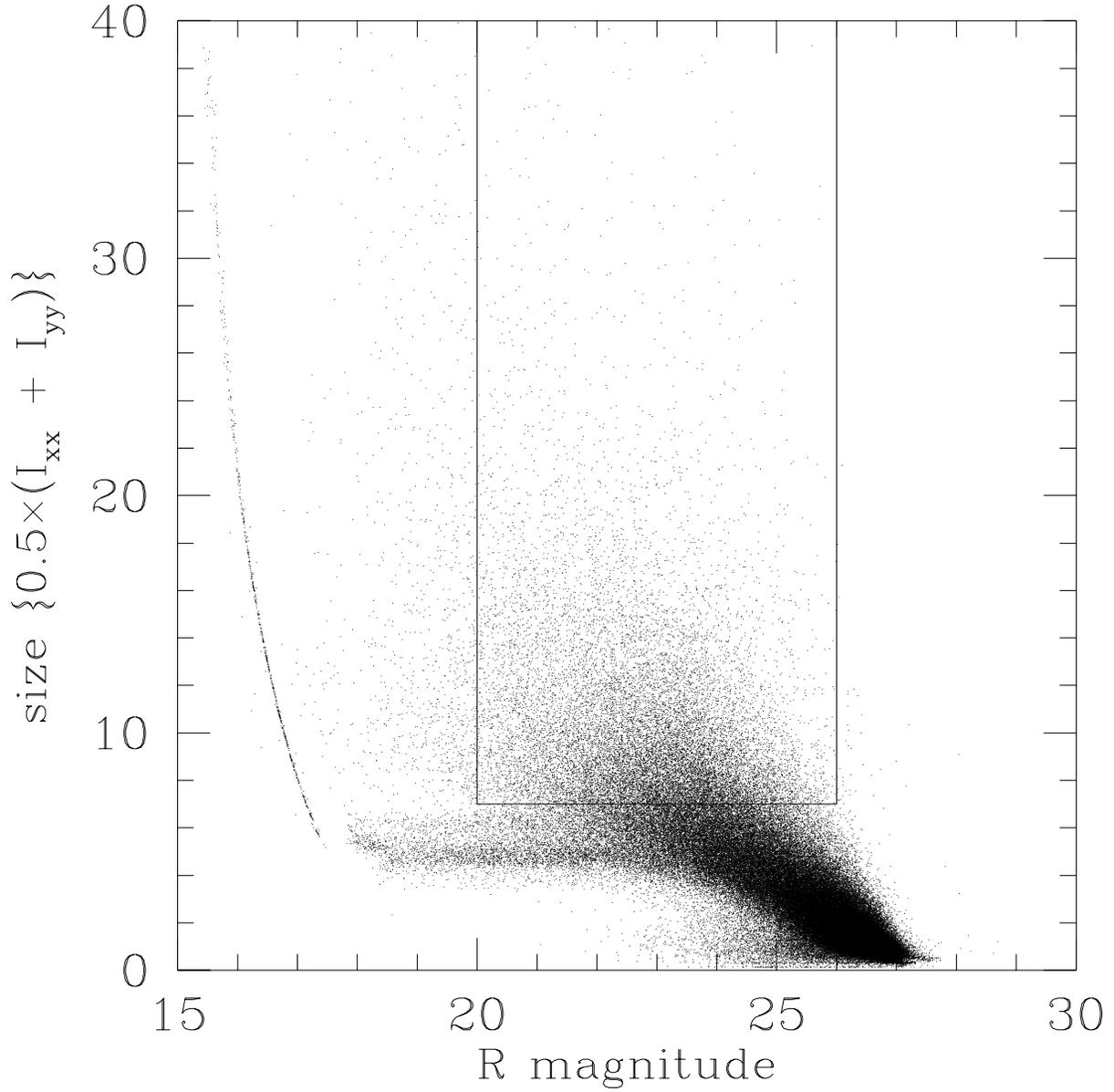} \caption{A plot of size versus $R$-magnitude for
well-defined objects in our sample. Potential background galaxies
were selected within the ``U'' shaped region shown for detection
of a weak lensing signal. \label{g_select}}
\end{figure}

\begin{figure}
\plotone{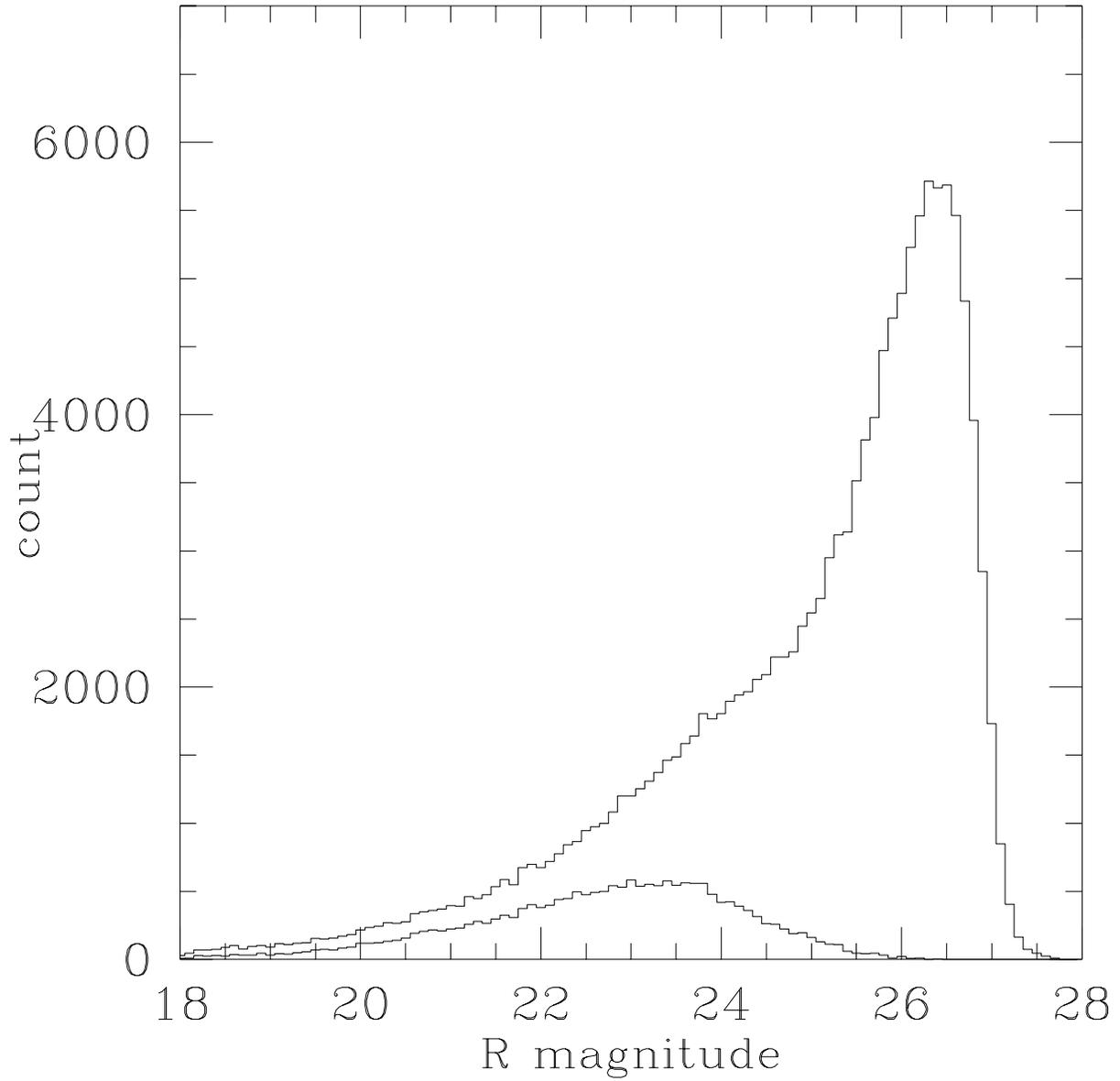} \caption{Two histograms indicating the number of
objects of a given $R$ magnitude.  The upper histogram shows all
objects from Fig.~\ref{g_select}, while the lower histogram shows
only those objects whose size, $0.5 \times (I_{xx} + I_{yy})$, is
greater than $7.0$. \label{Rselect} }
\end{figure}

Figure~\ref{g_select} shows a plot of size versus magnitude of the
objects whose shape {\tt{ellipto}} could determine.  Saturated
objects appear in the arc along the left hand side, while most
stars appear in the leftward arm.  We see in Fig.~\ref{Rselect}
that almost all of the objects that can be distinguished from the
stars by a size cut $0.5 \times (I_{xx} + I_{yy}) > 7$ have $R$
magnitudes between $18.0$ and $26.0$.

In principle, the color, $c = B_j - R$, of the cluster can be used
to eliminate foreground galaxies by selecting only those objects
whose color is more blue than the cluster.  However, preliminary
analysis showed that color cuts generally reduced the strength of
the weak lensing signal in this study.

The reason that color cuts reduced the measured shear in the
current study was two-fold.  First, because the seeing was
relatively poor in the $B_j$ images, approximately $10\%$ of the
objects in our sample did not have well determined $B_j$
magnitudes, even though they had well measured shapes in the $R$
image. Hence, any color cut excluded a substantial number of
objects in the background of the cluster whose shapes were well
measured. Second, while color cuts do eliminate most foreground
objects, they also eliminate some background objects that
contribute to the shear.

Since we can not rely on color selection, the primary object
selection criteria of importance will be the $R$ magnitude cut. We
present in this section and in our analysis two different $R$
selections.

It would appear that the conservative way to proceed would be to
make an $R$ magnitude cut above the ``expected'' $R$ magnitude of
cluster members at redshift $0.45$.  This would seem to ensure
that potential cluster alignment does not contaminate the weak
lensing signal.  Based on data from \citet{fuku}, cluster members
at $z=0.45$ have expected $K$-corrected $B_j$ magnitudes of $~23$,
and the expected $B_j-R$ color of such members is approximately
$2.7$.  (This agrees well with a sample of potential cluster
members hand selected from our image.) Therefore, an $R$ magnitude
cut of $R>22$ conservatively selects objects likely to fall behind
the cluster.

The more aggressive way to proceed is to choose a lower $R$
magnitude cut that maximizes the total detected gravitational
shear.  This is reasonable because, while the $R>22$ cut does
exclude most objects in the foreground of the cluster, it also
excludes objects {\emph{behind}} the cluster that do contribute to
the gravitational shear.

In Section 5, we estimate the redshift of each potential
background galaxy using statistics from the California Institute
of Technology Faint Galaxy Redshift Survey \citep{FGRS}. Applying
statistics drawn from that survey to our sample, we expect that
$78.7\,\%$ of objects whose $R$ magnitudes lie between $20 < R <
22$ fall behind the cluster, and overall, $90.1\,\%$ of all
objects with $20<R<26$ are expected to fall behind the cluster.
For the more restrictive $R$ magnitude cut at $R>22$, the overall
number of objects expected to fall behind the cluster is $93.6 \,
\%$, a slight increase from $90.1\,\%$ for the $20<R<26$ cut.
However, the $22<R<26$ cut excludes $22.2\,\%$ of all objects with
$20<R<26$ expected to fall behind the cluster, significantly
increasing the noise.

Our preliminary analysis determined that a simple $R$ magnitude
cut at $R>20$ was more effective at maintaining a high number of
background objects than color cuts or cuts based on the expected
cluster luminosity function. These findings are consistent with
the preliminary analysis of \citet{hossein}, who has analyzed DLS
deep field data for combinations of parameters that lead to the
most robust weak-lensing detections.

In the analysis sections of this paper, we present best fit model
parameters for both the $20<R<26$ and $22<R<26$ cuts with a size
restriction $0.5 \times (I_{xx} + I_{yy})$ greater than $7$.
However, we feel more confident in the $20<R<26$ cut as shown in
the U-shaped region of Fig.~\ref{g_select}. Using this constraint,
we obtained a well-measured background galaxy sample of $19796$
galaxies, or approximately $11$ galaxies per square arc~minute.

\subsection{Systematic errors and weak lensing detection}

Residual PSF anisotropy or inherent background galaxy ellipticity
correlations (independent of any lensing) would lead to systematic
errors in the measurement of the gravitational shear.  In this
subsection, we present evidence that any remaining PSF anisotropy
or inherent background galaxy ellipticity correlations are minimal
in strength compared to the weak lensing signal.

The combined $R$ band image of RXJ1347-1145 used in this study is
shown in Fig.~\ref{image}.  The cluster is located slightly
northeast of center in the frame.  Figure~\ref{fiatmap} shows the
two-dimensional mass map determined by the ellipticity
correlations in the $19796$ potential background galaxies for
$20<R<26$. This mass map has the same scale as the $R$ band image,
and was produced by {\tt{fiatmap}}, a program that determines a
two-dimensional mass profile from detected correlations in object
ellipticity \citep{wittman, FT}.

The color contrast mass map shows a strong (bright) weak lensing
signal at the location of the cluster. The field appears
relatively empty aside from a concentration of mass at the
location of the cluster.

Using the center of the cluster determined from the mass map, we
determine the average gravitational shear in radial annuli
centered on the cluster. This task is accomplished by determining,
for each object in a given radial annulus, the alignment of the
object to a circle centered on the cluster and then averaging
these values. For each annulus, we obtain an average tangential
shear, $\gamma_t$, (a measure of how aligned objects are to the
circle) and an estimated error in the tangential shear.

Figure~\ref{comb:fig} shows the tangential shear detected for
RXJ1347--1145 with $1\sigma$ error bars from analysis of the
$19796$ background galaxies with $20<R<26$. Also shown is the
contribution of the residual PSF anisotropy to the tangential
shear as determined by the stars. We see that the residual PSF
anisotropy is consistent with zero shear detection and
substantially weaker than the detection based on the potential
background galaxies.

In addition to measuring average tangential shear, we can measure
an average shear oriented at $45^\circ$ from tangent to the circle
centered on the cluster.  In weak lensing, this $45^\circ$ shear
should be consistent with zero shear detection at the large radii
of this study.  In dashed lines, Fig.~\ref{comb:fig} shows the
$45^\circ$ shear detected from the $19796$ background galaxies
with $20<R<26$. That the $45^\circ$ shear is consistent with zero
gives another indication that major systematic errors have been
controlled.

As a third test of systematic error, we show the results of an
$x-y$ scramble test in Fig.~\ref{scramble:fig}.  In this test, we
consider the quintuplet $(x^i,y^i,I^i_{xx}, I^i_{yy}, I^i_{xy})$
for the $i$th object used in the shear detection of
Fig.~\ref{comb:fig} and decouple the position $(x^i, y^i)$ from
the shape information $(I^i_{xx}, I^i_{yy}, I^i_{xy})$ by randomly
assigning shape information to positions, forming the quintuplets
$(x^i, y^i, I^{r}_{xx}, I^{r}_{yy}, I^{r}_{xy})$, where $r$
represents the same random integer for each moment.

Since weak lensing produces a position dependent shearing, the
quintuplet $(x^i,y^i,I^i_{xx}, I^i_{yy}, I^i_{xy})$ should result
in tangential shearing, but the scrambled quintuplet $(x^i, y^i,
I^{r}_{xx}, I^{r}_{yy}, I^{r}_{xy})$ should not. In
Fig.~\ref{scramble:fig}, we see that the $x-y$ scrambled
tangential shear is consistent with zero signal.

We statistically test the null hypothesis (that the true value of
$\gamma_t$ is zero) by computing a $\chi^2$ statistic assuming
that the error in the true value of $\gamma_t$ is equal to the
error in the measured $\gamma_t$.  This results in a $\chi^2$ of
$42.0$ for $15$ degrees of freedom for the sample with $20<R<26$.
The probability of obtaining a $\chi^2$ this large in the absence
of a weak lensing effect is less than $p = 0.0002$. The more
``conservative'' $R$ magnitude cut at $22<R<26$ yields $\chi^2 =
28.5$ for $15$ d.o.f., which corresponds to a probability of null
detection less than $p = 0.02$.

For comparison, a null hypothesis test for the $\chi^2$ statistic
for the $45^\circ$ shear shown in Fig.~\ref{comb:fig} results in a
$\chi^2$ of $16.6$ for $15$ degrees of freedom, while the
tangential shear measurement of the stars had a $\chi^2$ of
$26.6$.  The $x-y$ scramble test yields a $\chi^2$ of $11.4$ for
$15$ degrees of freedom, and is consistent with a null detection
at $p=0.724$.


\section{Mass reconstruction}

In gravitational lensing, the tangential shear, $\gamma_t$, at a
given projected radius $r$ in the lens plane is related to the
surface mass density, $\Sigma$, at that radius.  Defining $\kappa
= \Sigma / \Sigma_{crit}$, the ratio of the surface mass density
to the critical surface density for multiple lensing, we have

\begin{equation}
\gamma_t (r) = \bar{\kappa} (\le r) - \bar{\kappa} (r),
\label{shear}
\end{equation}

\noindent where $\bar{\kappa} (\le r)$ is the normalized, average
surface mass density interior to $r$ and $\bar{\kappa} (r)$ is the
mean density at $r$ \citep{miralda}.  For a given mass model that
determines $\Sigma$, Eq.~\ref{shear} allows us to relate the model
parameters to the observed shear, $\gamma_t$.

The critical surface density is given by \citep{ehlers}

\begin{equation}
\Sigma_{crit} = \frac{c^2}{4 \pi G} \frac{D_s}{D_l D_{ls}},
\label{sigma_crit}
\end{equation}

\noindent where $D_s$, $D_l$ and $D_{ls}$ are the angular-diameter
distances between the observer and source, observer and lens, and
lens and source, respectively.  Hence, $\Sigma_{crit}$ is both
redshift and cosmology dependent.

We will determine an average $\Sigma_{crit}$ for our sample by
using the known redshift of the cluster, $z=0.4509 \pm 0.003$
\citep{cohen-kneib}, and an average value of $\Sigma_{crit}$ for
each the many sources.

To determine the redshifts of the sources, we develop a redshift
to $R$ magnitude relation using the California Institute of
Technology Faint Galaxy Redshift Survey \citep{FGRS}. Dividing the
well measured objects in this survey into $0.50$-wide $R$
magnitude bins from $19.5<R<24.5$, we determined an initial
average value of $z$ for a given $R$ magnitude.  To remove
outliers from each bin, we excluded objects whose measured
redshift was more than one standard deviation from the bin
average, and then recomputed a final average redshift for the
given $R$ magnitude.

Plotting this average redshift of each bin against $R$ magnitude,
and using a weighted least squares method, we determine that $R$
is related to $z$ in a linear fashion

\begin{equation}
z=m \times R + b, \label{z-R_rel}
\end{equation}

\noindent where the parameters $m$ and $b$ are given by

\begin{eqnarray} m &=& 0.094 \pm 0.055, \nonumber \\ b &=& -1.5
\pm 1.2. \label{z-r_params} \end{eqnarray}

\noindent The slope-intercept parameter space is degenerate, so
the error bars on $m$ and $b$ from Eq.~\ref{z-r_params} are
linked.  Figure~\ref{ms:fit:fig} shows the binned data from the
CalTech Faint Galaxy Redshift Survey, the best fit line from
Eq.~\ref{z-R_rel} and two bounding curves corresponding to a
$68\%$ confidence interval.

The Faint Galaxy Redshift Survey does not contain a sufficient
number of galaxies fainter than $R>24.5$ to rule out a possible
selection bias.  Therefore, we posit that the linear fit found
from the $19.5<R<24.5$ region of the survey will apply to our
entire galaxy sample in Fig.~\ref{g_select}, even though our
sample contains objects fainter than $R>24.5$.

We note that any error introduced in the redshift by our linear
fit for objects with $R$ magnitude greater than $24.5$ will have a
minimal impact on the eventual determination of $\Sigma_{crit}$
for two reasons. First, from Fig.~\ref{Rselect}, we see that
relatively few objects will be used at $R>24.5$. Second, almost
all objects this faint will lie in the vicinity of $z \approx 1$
where the angular diameter distance flattens as a function of $z$.
Hence, even relatively large errors in $z$ around $z \approx 1$ do
not propagate strongly into errors in $D_s$, $D_{ls}$ or
$\Sigma_{crit}$.

Applying Eq.~\ref{z-R_rel} to our sample with $20<R<26$, we
determine that our sample has an average redshift, ${\bar{z}} =
0.74 \pm 0.13$, and that the average critical surface density is
$\Sigma_{crit} = 8200^{+3100}_{-1400} h M_\odot$ pc$^{-2}$ for an
$\Omega_m = 1$, $\Omega_\Lambda = 0$ cosmology with $H = 50$ km
s$^{-1}$ Mpc$^{-1}$.  (Here, the upper and lower error limits are
estimated using the bounding curves from Fig.~\ref{ms:fit:fig}.)
For an $\Omega_m = 0.3$, $\Omega_\Lambda = 0.7$ cosmology, the
average $\Sigma_{crit}$ is given by $6000^{+2300}_{-1000} h
M_\odot$ pc$^{-2}$.  For the sample with $22<R<26$, the values for
$\Sigma_{crit}$ are $8000^{+3200}_{-600} h M_\odot$ pc$^{-2}$ and
$6500^{+2100}_{-1000} h M_\odot$ pc$^{-2}$ for $\Omega_m = 1$,
$\Omega_\Lambda = 0$ and $\Omega_m = 0.3$, $\Omega_\Lambda = 0.7$
cosmologies, respectively.

In principle, the critical density varies as a function of angular
radius because the magnification by the lens of background sources
makes sources closer to the center of the lens appear relatively
brighter than sources at a greater angular radius.  This increase
in brightness causes us to underestimate the redshift of the
source and overestimate the critical density.

\citet{FT} used Monte Carlo simulations of background galaxies
magnified by an isothermal lens with a velocity dispersion of
$1460$ km s$^{-1}$ to determine a plot of $\Sigma_{crit}$ versus
radius.  Figure~7 of their paper shows that the observed shear of
their study's innermost data points were affected by the variation
in $\Sigma_{crit}$. However, accounting for this variation did not
lead to a significant change in model fit, primarily because the
innermost angular bins have the fewest data points and hence the
highest uncertainty.

In the current wide-field study, the innermost radial bin falls
over the range in which Fig.~7 of \citet{FT} shows a significant
variation in $\Sigma_{crit}$.  However, the remaining radial bins
fall into a region where the variation of $\Sigma_{crit}$ is
within the error introduced by using the fit of Eq.~\ref{z-R_rel}.

Therefore, in this wide-field study, we will use a constant value
of $\Sigma_{crit}$ as a function of angular radius for a given
cosmological model.  We feel justified in doing so because the
current wide-field study covers almost twice the radial size of
\citet{FT} and hence is less susceptible to any biasing introduced
by errors from the innermost data point. These errors are
 small because the weighted $\chi^2$ algorithms used to
fit mass models discount data points with large errors.


\section{Mass models}

In this section, we apply two mass models to our observed shear of
RXJ1347-1145: a singular isothermal sphere model (SIS) and the
Navarro, Frenk, and White (NFW) model \citep{nfw}.  Throughout
this section, we assume that the Hubble constant has a value of $H
= 50$ km s$^{-1}$ Mpc$^{-1}$.  For each model, we consider an
$\Omega_m = 0.3$, $\Omega_\Lambda = 0.7$ cosmology, and briefly
cite the best fit parameters for the $\Omega_m = 1$,
$\Omega_\Lambda = 0$ cosmology in order to compare with previous
studies of this cluster. In~\ref{om:sec}, we summarize the cluster
masses inferred from the two models.

\subsection{SIS model}

In the SIS model, the surface density is parameterized by the velocity
dispersion, $\sigma_v$:

\begin{equation}
\Sigma (r) = \frac{\sigma_v^2}{2G} \frac{1}{r}.
\label{SIS}
\end{equation}

\noindent Here, $r$ is the projected radius in meters in the lens
plane from the center of the cluster. In the SIS model,
$\bar{\Sigma}(r) = 2\Sigma(r)$, so by Eq.~\ref{shear}, the
tangential shear is related to the velocity dispersion by

\begin{eqnarray}
\gamma_t (r) &=&  \bar{\kappa} (\le r) - \bar{\kappa} (r)
\nonumber \\ ~&=& \frac{\Sigma (r)}{\Sigma_{crit}} = \frac{1}{\Sigma_{crit}}
\left( \frac{\sigma_v^2}{2\, G} \right) \frac{1}{r}. \label{SIS_shear}
\end{eqnarray}

\noindent Using a standard linear regression fitting algorithm
that minimizes $\chi^2$, we determine that the velocity dispersion
predicted by our observed shear in an $\Omega_m = 0.3$,
$\Omega_\Lambda = 0.7$ cosmology is $\sigma_v = 1400$ km s$^{-1}$
 for the $20<R<26$ magnitude cut (minimum $\chi^2=14.8$ for $14$
 d.o.f.), and $1550$ km s$^{-1}$ for the $22<R<26$ magnitude cut
 (minimum $\chi^2=12.3$ for $14$ d.o.f.)

Two independent factors contribute to error in the inferred
velocity dispersion. First, the process of minimizing $\chi^2$
carries with it a random statistical error that can be estimated
using the $\Delta \chi^2$ argument of \citet{nrc}.  We estimate
that these statistical errors contribute to uncertainty in the
velocity dispersion of $\sigma_v = 1400^{+130}_{-140}$ km s$^{-1}$
for a $68\%$ confidence interval with the $20<R<26$ cut and
$\sigma_v = 1550^{+180}_{-210}$ km s$^{-1}$ for the $22<R<26$ cut.
Independently, the errors bars on the lensing critical density,
contribute to uncertainty in $\sigma_v$ as $\sigma_v =
1400^{+240}_{-120}$ km s$^{-1}$ for $20<R<26$ and
$1550^{+230}_{-130}$ km s$^{-1}$ for $22<R<26$.

Assuming the velocity dispersion of $\sigma_v =
1400^{+130}_{-140}$ km s$^{-1}$, the SIS model corresponds to a
total integrated mass of $1.4 \, \pm \, 0.3 \, \times 10^{15} \,
M_\odot$ within a radius of $1.0$ Mpc.

For an $\Omega_m = 1.0$, $\Omega_\Lambda = 0$ cosmology, the best
fit velocity dispersions are $1500^{+140}_{-150}$ km s$^{-1}$
(statistical) and $1500^{+260}_{-130}$ km s$^{-1}$ (error in
critical density) for $20<R<26$ and $1580^{+180}_{-210}$ km
s$^{-1}$ (statistical) and $1580^{+290}_{-60}$ km s$^{-1}$
(critical density) for $22<R<26$.

\subsection{NFW model}

The NFW model depends on a concentration parameter $c$, and a scale
radius $r_s$.  The scale radius is related to the virial radius
$r_{200} = c\,r_s$, which is the radius inside which the mass density of
the halo is equal to $200\rho_c$, where $\rho_c = \frac{3H^2(z)}{8\pi G}$
is the critical density of the universe at redshift $z$.

From Eq.~\ref{shear}, we have that

\begin{equation} \gamma_t (r) = \frac{\bar{\Sigma}_{NFW} (\le r) - \Sigma_{NFW} (r)}
{\Sigma_{crit}}, \end{equation}

\noindent and \citet{wright} derive the relevant expressions for
$\bar{\Sigma}_{NFW} (\le r)$ and $\Sigma_{NFW}(r)$.  Because our
observational data is specified in terms of annular bins with
inner and outer radii $r_i$ and $r_o$, we must derive expressions
for the average density interior to the annulus and the average
density in the annulus. The average density {\emph{in}} the
annulus is given by

\begin{equation}
\Sigma (r) = \bar{\Sigma} (r_o) \frac{r_o^2}{r_o^2 - r_i^2} -
\bar{\Sigma} (r_i) \frac{r_i^2}{r_o^2 - r_i^2}, \label{sig_in}
\end{equation}

\noindent while the average density {\emph{interior}} to the
annulus is given by

\begin{equation}
\bar{\Sigma} (r) = \bar{\Sigma} (r_o) \frac{r_o^2}{r_o^2 + r_i^2} +
\bar{\Sigma} (r_i) \frac{r_i^2}{r_o^2 + r_i^2}. \label{sig_inside}
\end{equation}

\noindent Hence, our observed tangential shear, $\gamma_t$, for a
given annulus of inner radius $r_i$ and outer radius $r_o$ is given by

\begin{equation}
\gamma_t = \frac{2}{\Sigma_{crit}} \left( \frac{r_o^2 \, r_i^2}
{r_o^4 - r_i^4} \right) \left( \bar{\Sigma} (r_i) - \bar{\Sigma} (r_o)
\right). \label{NFW_shear}
\end{equation}

We use Eq.~\ref{NFW_shear} with the expressions for $\bar{\Sigma}$
from \citet{wright} to determine the best fit parameters of the
NFW model. As the NFW model is highly non-linear, we use the
\citet{nrc} implementation of the Levenberg-Marquardt method to
fit for $c$ and $r_s$, minimizing $\chi^2$.

For an $\Omega_m = 0.3$, $\Omega_\Lambda = 0.7$ cosmology, we find
that $c = 15$, and $r_s = 0.23$ Mpc, or $r_{200} = 3.5$ Mpc for
the $20<R<26$ cut with a minimum $\chi^2 = 15.3$ for thirteen
d.o.f. For $22<R<26$, we obtain $c = 6.8$, and $r_s = 0.69$ Mpc,
or $r_{200} = 4.7$ Mpc with $\chi^2 = 10.2$. Figure~\ref{fits}
shows the best fit NFW model with the observed gravitational shear
for $20<R<26$.

As for the SIS models, we estimate the statistical errors in the
$\chi^2$ minimization and the error propagated from error in
$\Sigma_{crit}$. In the case of the NFW models, the error in
$\Sigma_{crit}$ introduces a small error in the concentration
parameter, $c = 15^{+2}_{-1}$ for $20<R<26$ and $c =
6.8^{+0.9}_{-0.5}$ for $22<R<26$ and less than $1\%$ variation in
$r_s$ for both cuts.

Figure~\ref{NFW_err} shows the $68\, \%$ confidence region in the
$c$--$r_{200}$ plane for the $20<R<26$ cut, yielding error bars
$r_{200} = 3.5^{+0.8}_{-0.2}$ Mpc with $c = 15^{+64}_{-10}$ for a
NFW model in an $\Omega_m = 0.3$, $\Omega_\Lambda = 0.7$
cosmology. The concentration parameter, $c$ is not well
constrained. (The confidence region for $22<R<26$ is similar.)

We estimate error bars on the total integrated mass using the
upper left and lower right projections of the confidence interval
onto the $c$--$r_{200}$ axes. The lack of constraint on $c$ yields
large error bars on the total integrated mass: $2.7^{+2.6}_{-1.4}
\times 10^{15} M_\odot$ within $r_{200}$ and $1.4^{+0.5}_{-0.4}
\times 10^{15} M_\odot$ within $1.0$ Mpc in the $\Omega_m=0.3$,
$\Omega_\Lambda = 0.7$ cosmology for the $20<R<26$ cut.

The best fit NFW parameters for an $\Omega_m = 1$, $\Omega_\Lambda
= 0$ cosmology are $c = 11.5^{+1.6}_{-0.8}$ (error from
$\Sigma_{crit}$), $r_s = 0.193$ Mpc, or $r_{200} = 2.22$ Mpc with
$\chi^2 = 15.3$ for thirteen d.o.f. for $20<R<26$ and $c =
4.8^{+0.7}_{-0.2}$ (error from $\Sigma_{crit}$), $r_s = 0.584$
Mpc, or $r_{200} = 2.78$ Mpc with $\chi^2 = 1.2$ for thirteen
d.o.f. for $22<R<26$. Again, the concentration parameter, $c$ is
not well constrained by the statistical fit, as is shown in
Fig.~\ref{NFW_err}.

\subsection{Observed Mass}\label{om:sec}

Table~\ref{model:table} summarizes the inferred mass interior to
$1$~Mpc of the cluster in SIS and NFW models for two cosmological
models for the $20<R<26$ cut, while Table~\ref{model2:table}
presents the same information for the $22<R<26$ cut. We see that
within a cosmological model, the NFW and SIS inferred masses
generally agree within $1\sigma$ error bars, and that the choice
of cosmological model makes a small difference in inferred mass.
We also note that the $22<R<26$ cut generally yields a higher
total integrated mass than the $20<R<26$ cut.

Figure~\ref{mass:fig} plots an observed measure of the mass
densitometry of RXJ1347--1145 computed by the relation

\begin{equation} \delta \bar \kappa (r_i) = \bar \kappa (r<r_i) - \bar \kappa
(r_i<r<r_o), \label{densitometry} \end{equation}

\noindent introduced by~\citet{fahlman}, where $r_o$ is the outer
radius of the outermost annulus.  Densitometry measures from SIS
and NFW models in an $\Omega_m = 0.3$, $\Omega_\Lambda = 0.7$
cosmology are shown assuming model parameters from best-fits to
the gravitational shear data.  The error bars are statistical
error bars computed from the number of galaxies used in the
calculation only.

Because the observed densitometry measure uses all of the objects
between $r_i$ and $r_o$ to determine $\delta \bar\kappa$ at $r_i$,
the values of the observed densitometry measure in
Fig.~\ref{mass:fig} are not independent. In fact, small changes in
values at the outer annuli give rise to large changes in the
densitometry value at inner annuli, so that the actual uncertainty
in the innermost annulus is quite large. Thus, although neither
the SIS or NFW model fit the actual value of the innermost point,
both models are adequate fits to the total profile and yield
similar total integrated masses.


\section{Luminosity}

We present a rough estimate of the light distribution of the
cluster drawn from the $R$ band image in Fig.~\ref{lum:fig}. This
estimate is limited by the difficulty of accurate background
subtraction due to the lack of a high quality $B_j$ image that can
be used for color selection.  This luminosity estimate is also
limited by the compact nature of RXJ1347 and the relatively low
angular resolution of the central region in this study.  In this
section, all measured values are the \verb"SExtractor" values for
the combined $R$ band image.

We begin by selecting all non-saturated objects detected in the
$R$ band image by \verb"SExtractor" with a minimal lower limit
size cut $0.5\times(I_{xx}+I_{yy})>2$ and deleting spuriously
detected objects in regions near stars.  Because we can not make a
color or redshift selection, we {\emph{assumed}} that all objects
are at the cluster redshift, and determined a total flux,
$\textit{f}$, in radial annuli centered on the cluster. From the
total flux in each annulus, we determined a total absolute
magnitude $M$, applying a $K$ correction of $0.625$ from
\citet{fuku}. A luminosity relative to the solar luminosity in
each annulus was then computed from this total magnitude.

To compare this light distribution with the mass distribution
determined by the weak lensing, we first determined that the
luminosity per unit area is well fit by a general SIS profile.
Since this is the case, we can assume that the luminosity per unit
area is linearly related to the projected SIS mass density, or
that

\begin{equation} L_R / A = a + b \times \Sigma (r), \label{L/M}
\end{equation}

\noindent where the SIS model for mass density $\Sigma$ is given
by the best fit parameters found from the weak lensing, or
$\sigma_v = 1400^{+130}_{-140}$~km~s$^{-1}$ for the $\Omega_m =
0.3$, $\Omega_\Lambda = 0.7$ cosmology.  We compute $\Sigma (r)$
in units of solar masses per unit area and find the parameters $a$
and $b$ by minimization of

\begin{equation} \chi^2 = \left( \frac{L_R/A - (a + b \times
\Sigma (r))}{\delta \Sigma (r)} \right)^2 \label{chi_min}
\end{equation}

\noindent with

\[ \delta \Sigma (r) = \frac{b \, \sigma_v \, \delta \sigma_v}{G \, r}. \]

Minimizing $\chi^2$ from Eq.~\ref{chi_min} has the advantage of
finding both the approximate ``background'' luminosity and mass to
light ratio simultaneously.  Formally, the denominator of $\chi^2$
should also include a term for uncertainty in $L_R/A$. We neglect
this uncertainty because we are unable to determine its value
without information about the background redshift distribution,
although it is reasonable that it is approximately constant for
all radial bins.

The best fit parameters, $a$ and $b$, for seven evenly spaced
annuli out to approximately $2.5$~Mpc are
$a=1.9$~$L_{R\odot}$~pc$^{-2}$ and $b = 0.011$~$L_{R\odot} /
M_\odot$ with $\chi^2 = 10$ for five degrees of freedom.  This
implies that M/$L_R$ $=90~M_\odot / L_{R\odot}$ with $H = 50$ km
s$^{-1}$ Mpc$^{-1}$. Figure~\ref{lum:fig} plots the luminosity per
unit area and best fit scaled mass density (from Eq.~\ref{L/M})
along with $1\sigma$ scaled mass density curves.

Since we do not have error estimates on the luminosity data, we
determine only the statistical uncertainty present in the fit
using the $\Delta \chi^2$ argument of \citet{nrc}.  This yields an
error ellipse whose projections indicate that the $68\%$
confidence interval for the mass to light ratio is M/$L_R$ $=91
\pm 20 ~M_\odot / L_{R\odot}$.


\section{Discussion}

In this section, we discuss the two $R$ magnitude cuts and compare
our results with the previous weak lensing results of \citet{FT},
the recent X-ray studies of \citet{allen2002}, and the recent
optical studies of \citet{cohen-kneib}.   Our results indicate
that there were not systematic problems in the previous weak
lensing studies.

\subsection{Object selection criteria}

Because our $B_j$ imaging was poor, color selection criteria
dramatically reduced the significance of our weak lensing
detection by eliminating objects with well measured shapes that
were undetected in the $B_j$ image.  Thus, we were forced to
employ only size and $R$ magnitude cuts to restrict our image. A
size cut $(0.5\times\{I_{xx}+I_{yy}\}>7)$ was used to eliminate
poorly resolved objects and stars, while the $R$ magnitude cut
essentially was used to control foreground objects (see
Fig.~\ref{g_select}).

In principle, it is very important to eliminate foreground and
cluster objects from the catalog before analyzing gravitationally
induced shears for two reasons.  First, foreground objects weaken
the signal because their shapes are not correlated with the weak
lensing shear. Second, one might conjecture that cluster members
align in some correlated way and unduly influence the
gravitational shear detection if they are included.

In practice, we find that ``restrictive'' object selection
criteria do not enhance the weak lensing detection.  In this
study, the conservative $22<R<26$ yielded a substantially lower
total $\chi^2$ relative to the null detection ($\chi^2=28.5$, $15$
d.o.f.) than the $20<R<26$ cut ($\chi^2=42.0$, $15$ d.o.f.).

The $R$ magnitude cut at $R>22$ is reasonable based on the
expected cluster $R$ magnitude of $R\approx 20$ at $z=0.45$.  This
cut successfully eliminates most foreground and cluster members
from further analysis.  This would apparently lead to a strong
detection of gravitational shear by reducing the noise.

However, a cut at $R>22$ also eliminates a significant number of
galaxies in the background of the cluster that are gravitationally
lensed.  The effect of removing these lensed objects from analysis
weakens the signal much more than the signal is enhanced by
removing the noisy foreground, and so the weak lensing detection
is more significant with a less restrictive cut at the expected
cluster $R$ magnitude.

\begin{figure}
\plotone{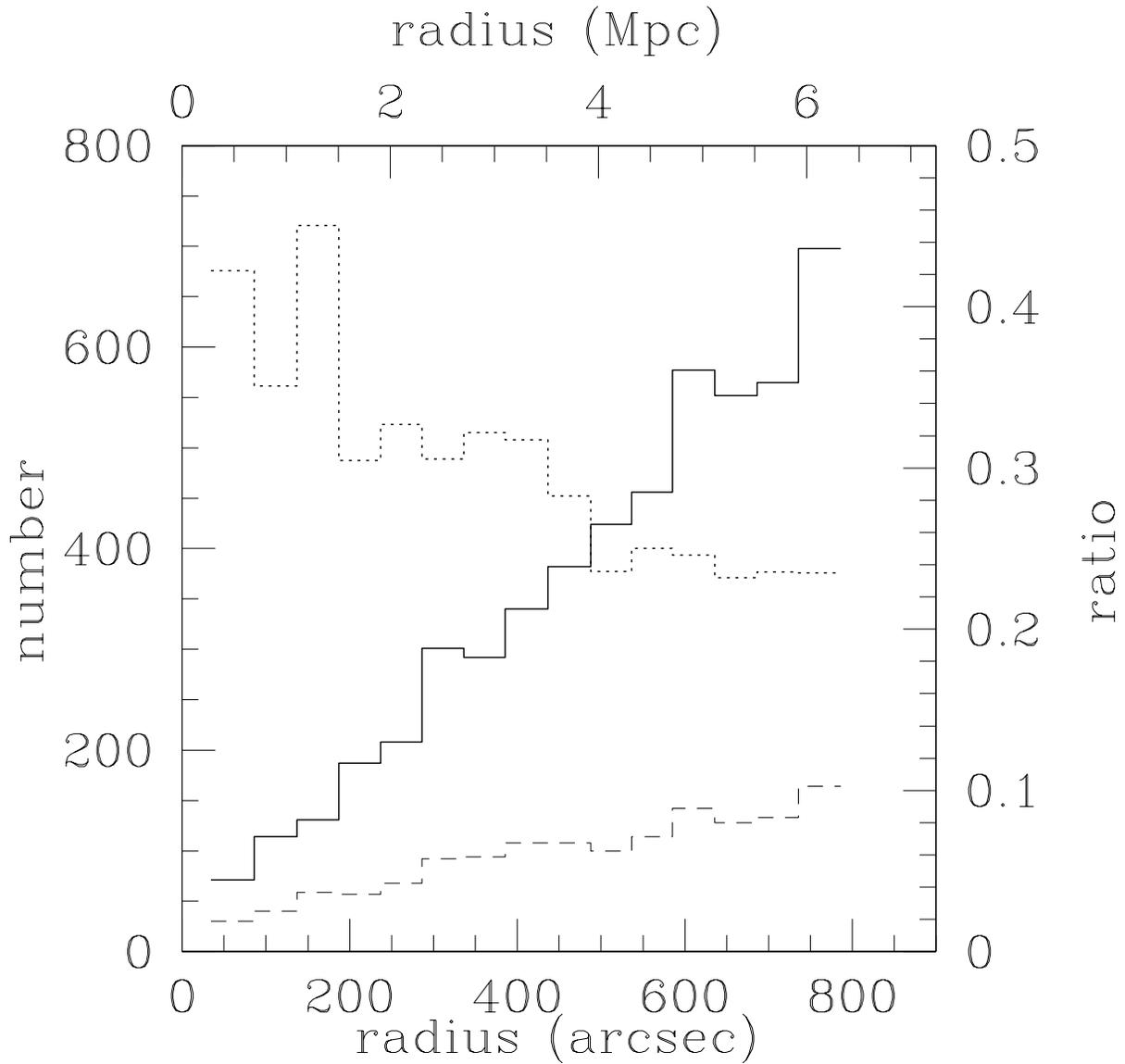} \caption{Three histograms indicating the total
number of objects (solid), the number of objects with $20<R<22$
(long dashed), and ratio (short dashed) in radial bins. That the
ratio of the numbers of bright objects and all objects is
relatively constant indicates that the inclusion of objects with
$20<R<22$ does not significantly contaminate the sample.}
\label{rnums}
\end{figure}

\begin{figure}
\plotone{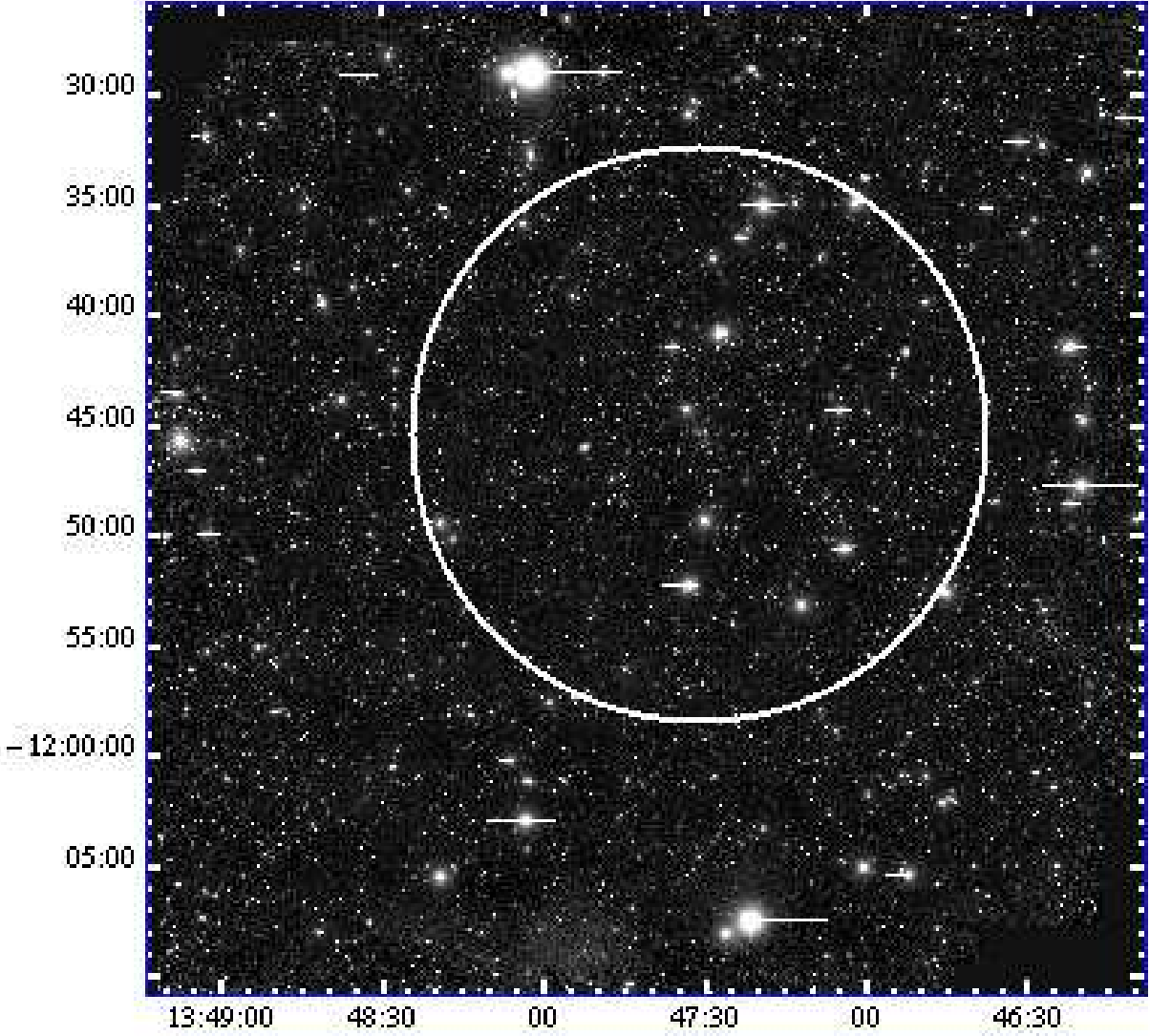} \caption{The combined $R$ band image of field
around RXJ1347-1145 used in this study.  The circle, centered on
the cluster at an RA of approximately $13$:$47.30$ and Dec of
$-11$:$45$, indicates the region of the image analyzed for weak
lensing signals. \label{image} }
\end{figure}

We do not believe that inclusion of objects with $20<R<22$
significantly contaminates the sample or biases the weak lensing
detection.  In Fig.~\ref{rnums}, we see that objects in the
innermost radial bins are slightly brighter than in the outer
bins. However, if we consider the over-density of objects with
$20<R<22$ in the ``cluster'' region interior to an angular radius
of $r<400''$, we estimate that less than $100$ additional objects
have been added to the catalog of over $1300$ total objects with
$r<400''$.

Further, we observe that the mass inferred from the best fit
parameters of SIS and NFW models for the sample of objects with
$22<R<26$ is only slightly higher than that from objects with
$20<R<26$.  In fact, the statistical error bars have significant
overlap for the two $R$ magnitude cuts.

We believe that the results from the $20<R<26$ magnitude cut
represent a better measurement of weak gravitational lensing than
the $22<R<26$ magnitude cuts because the significance of detection
is much higher.  However, the best fit model parameters are
similar, and the basic conclusions of the paper are unchanged for
either $R$ magnitude cut.

\subsection{Previous weak lensing observation}

The general results of this paper are in agreement with the
results of \citet{FT}, who report that the total integrated mass
within $r<1$ Mpc is $1.7 \pm 0.4 \times 10^{15} M_\odot$, and that
the velocity dispersion is $\sigma_v = 1500 \pm 160$ km s$^{-1}$
for SIS model assuming an isotropic velocity dispersion and an
$\Omega_m = 1$ cosmology. These values are consistent with our
results for the same model.

As is seen in Fig.~\ref{scramble:fig}, the wide-field weak lensing
observations presented here confirm the detection of \citet{FT},
which was limited to an approximately $400''$ radius. The current
wide-field study continues to detect significant gravitational
shear at $r>400''$; however, the impact this shear has on the
model parameters in SIS and NFW models is minimal.

The mass to light ratio of the cluster measured in this study
agrees relatively well with \citet{FT}, who find values of M/L$_R$
$= 75\,\pm 20$~M$_\odot / L_{R\odot}$.  This is quite remarkable
considering the difference in methods and difficulty of background
subtraction in the current study.

That the current weak lensing study obtains results in agreement
with the results of \citet{FT} gives a strong indication that the
concerns regarding the use of smaller images in determining
properties from lensing studies were unfounded. The large field of
view of this study and extremely robust weak lensing detection
strongly confirms the results of \citet{FT}.

\subsection{Recent X-ray observations}

Our results are also in good agreement with the recent Chandra
X-ray observations of \citet{allen2002}. Using an NFW model,
\citet{allen2002} determine a best-fit scale radius of $r_s
=0.40^{+0.24}_{-0.12}$ Mpc with a concentration parameter of
$c=5.87^{+1.35}_{-1.44}$.  These values correspond to a virial
radius $r_{200} = 2.35^{+0.49}_{-0.33}$ Mpc and an integrated mass
within the virial radius of $M_{200} = 2.29^{+1.74}_{-0.82} \times
10^{15} M_\odot$.

The current weak lensing observations find a similar virial
radius, $r_{200} = 2.22$ Mpc, and a consistent mass $M_{200} =
2.3^{+1.9}_{-1.1} \times 10^{15} M_\odot$ for the same
cosmological model. Figure~\ref{NFW_err} shows the $68 \, \%$
confidence region for the best fit parameters $r_s$ and $c$ for
the current weak lensing study with $20<R<26$ and the location of
the parameters found by \citet{allen2002}.  The recent Chandra
results are also consistent with a $22<R<26$ magnitude cut.

\citet{allen2002} report that a SIS model does not provide a good
fit to the Chandra observations.  Our observations are more
tolerant of a SIS model because we do not have lensing
observations close to the center of the cluster, a region that is
well sampled and integral for the X-ray study's conclusions.

\subsection{Recent spectroscopic observations}

This weak lensing study is not consistent with the results of the
most recent optical velocity dispersion study by
\citet{cohen-kneib}.  \citet{cohen-kneib} report a velocity
dispersion of $\sigma_v = 910\pm 130$ km s$^{- 1}$, corresponding
to an integrated mass of $6.0 \times 10^{14} M_\odot$ within $1.0$
Mpc. This predicted mass is approximately $38\,\%$ of the mass
estimated by the current weak lensing observations, and is
inconsistent with our results at the $5\sigma$ level.

\citet{cohen-kneib} suggest that their results may be artificially
low if RXJ1347-1145 consists of two clusters caught in the act of
merging in a direction perpendicular to the line of sight.  This
flow would have a minimal effect on the wide-field lensing signal
analyzed in this study because the wide-field imaging does not
have sufficient resolution to examine the interior region of the
cluster.


\section{Conclusions}

In conclusion, the findings of this paper confirm the earlier weak
lensing study of \citet{FT} and are consistent with the recent X-
ray observations of \citet{allen2002} and \citet{ettori}.  From
these findings, we conclude that there were no substantial
systematic errors introduced in the older lensing study of this
system by \citet{FT} through the relatively small field of view.
Confirming the older weak lensing observations of \citet{FT} using
a wide-field observation gives substantial evidence that weak
lensing may not be significantly biased by relatively small fields
of view.

This study's results are in substantial disagreement with the
direct optical velocity dispersion measurements of
\citet{cohen-kneib}. As the X-ray and lensing results now tend to
cluster around a velocity dispersion of approximately $1500$ km
s$^{-1}$, it is unlikely that the $910$ km s$^{-1}$ measurement of
Cohen and Kneib is correct.  \citet{cohen-kneib}, recognizing this
difference suggest that a possible merger within the cluster
perpendicular to the line of sight could account for the observed
difference.

In the current study, we did not attempt to measure weak lensing
signals interior to the two arc candidates located approximately
$35''$ from the center of the cluster.  Because the original goal
of this study was to test the small field of view of \citet{FT},
the wide-field image used in this study does not have enough
resolution or depth to conclusively measure weak lensing signals
in the central region of the RXJ1347-1145.

For this reason, we can not attempt to locate substructure within
the cluster to test Cohen and Kneib's assertion in this study or
compare our results to recent strong lensing observations. In
addition, we suspect that it is the X-ray observation's greater
sensitivity to the central region of RXJ1347-1145 that accounts
for the inability of \citet{allen2002} to find a good SIS fit.

\begin{figure}
\plotone{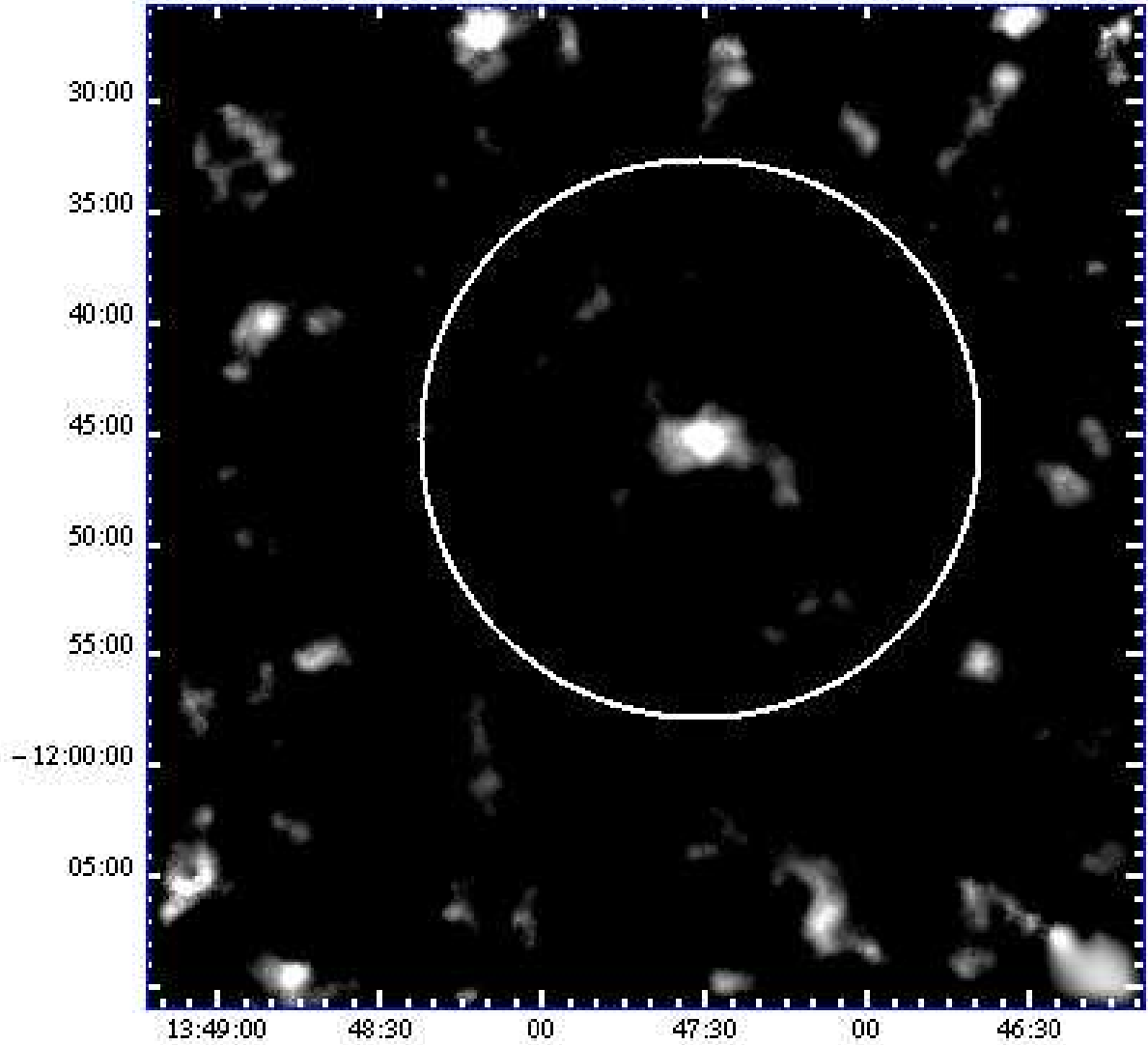} \caption{A projected two-dimensional mass map
for RXJ1347-1145 over a $43' \, \times 43'$ field of view. The
bright, compact, central object is the cluster.  The circle
indicates the region of the image analyzed for weak lensing
signals.  The relative lack of other substantial regions of mass
in the middle of the image indicates a significant weak lensing
detection about the center of RXJ1347-1145.  The features detected
near the edges of the image are artifacts of a reduced amount of
data at the boundaries. \label{fiatmap}}
\end{figure}

\begin{figure}
\plotone{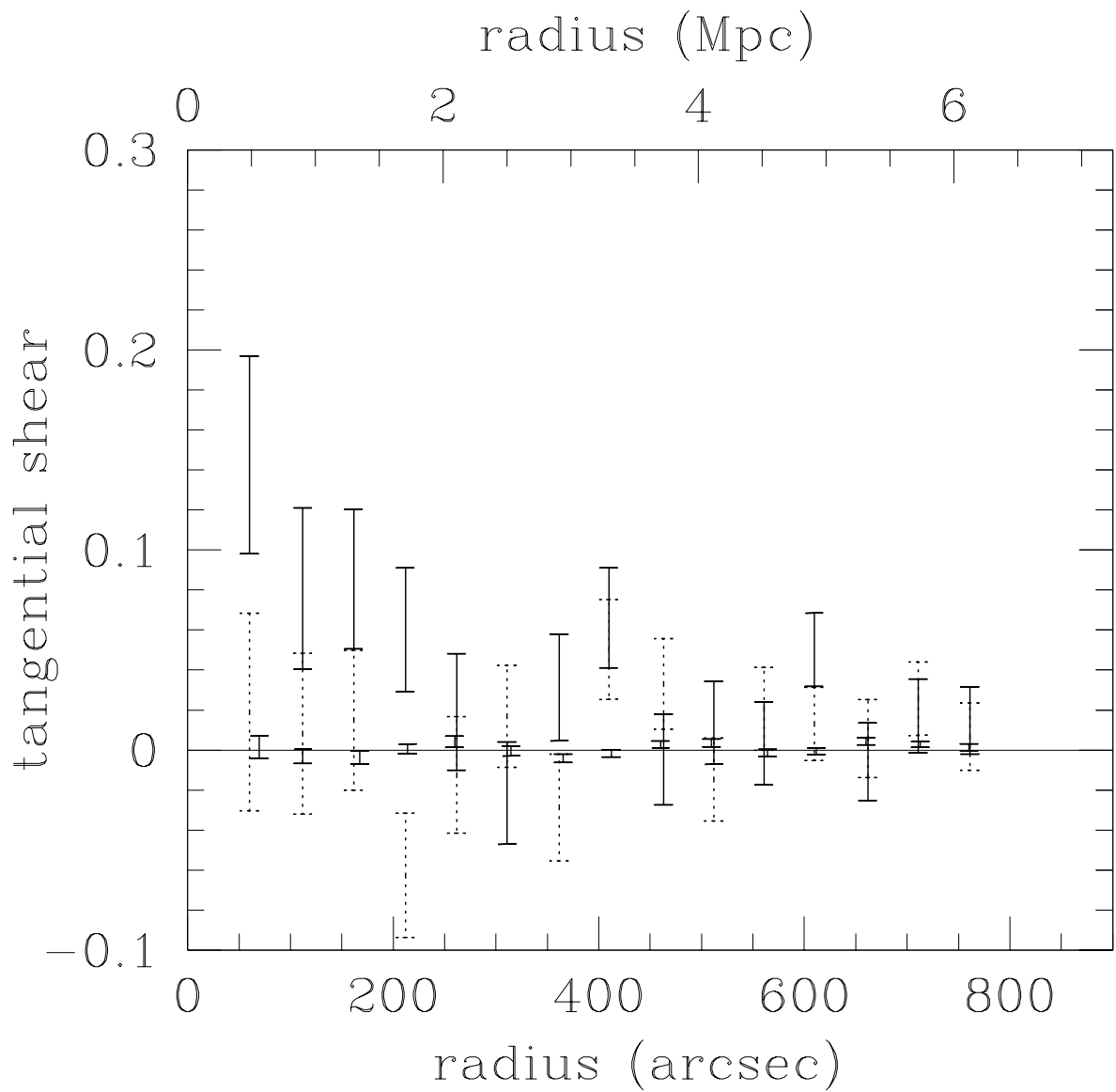} \caption{The tangential shear as a function of
angular radius due to weak lensing of background galaxies near
RXJ1347-1145 with $1\sigma$ error bars. Also shown is an estimate
of the contribution to the detected shear due to residual PSF
anisotropy as determined by analysis of stars (small solid errors
bars) and the $45^\circ$ shear as a function radius (large dashed
error bars). The Mpc scale is the radial distance in the lens
plane for an $\Omega_m = 0.3$, $\Omega_\Lambda = 0.7$ cosmology.
\label{comb:fig}}
\end{figure}

\begin{figure}
\plotone{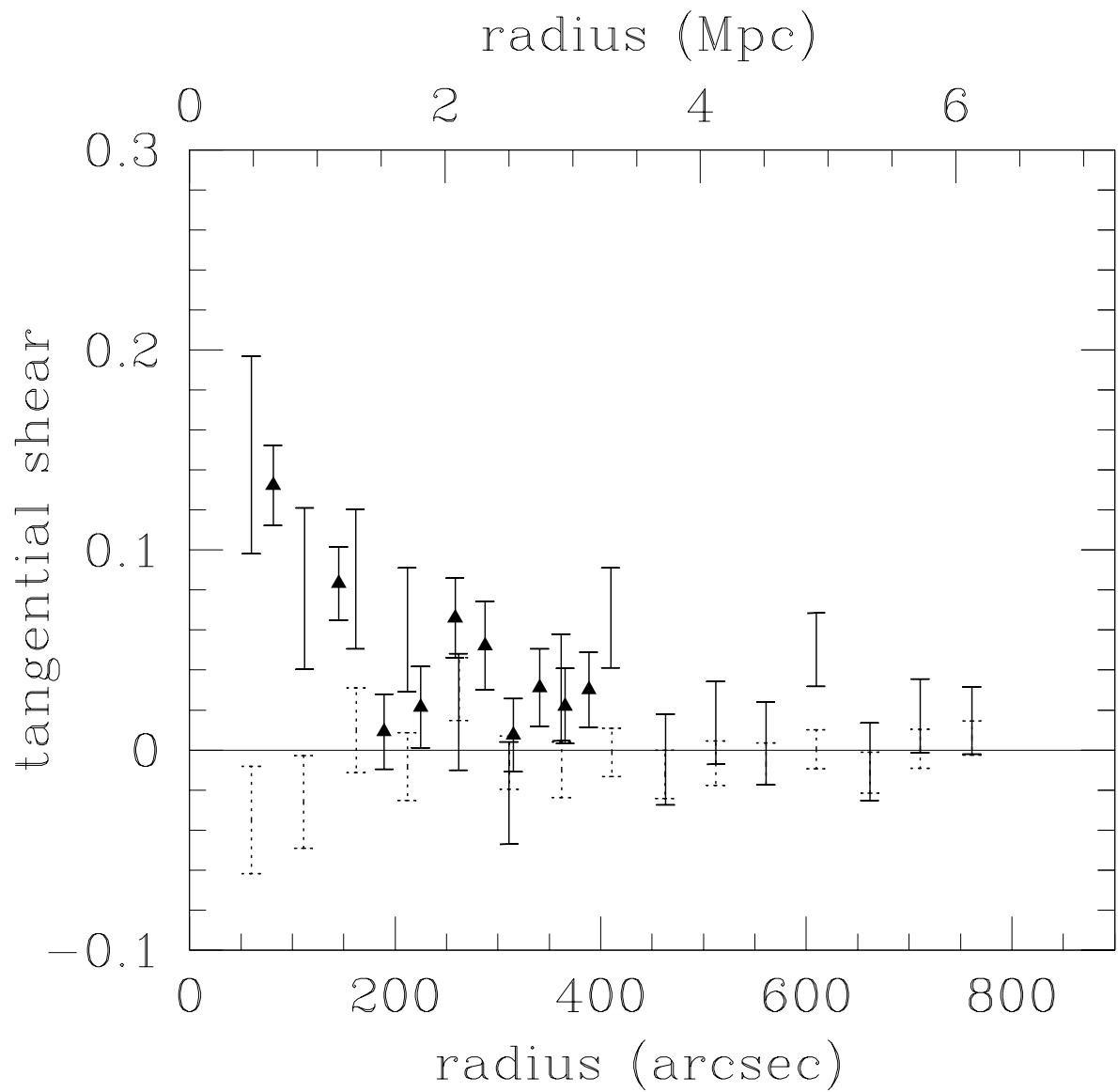} \caption{The tangential shear as a function of
angular radius with $1\sigma$ error bars is shown with the results
of an $x-y$ scramble test (dashed error bars).
 The $x-y$ scramble test is a measurement of systemic errors.
 Also shown are the data from an earlier weak lensing detection
 (triangles).
 The Mpc scale is the radial distance in the lens plane for an
$\Omega_m = 0.3$, $\Omega_\Lambda = 0.7$ cosmology.
\label{scramble:fig}}
\end{figure}

\begin{figure}
\plotone{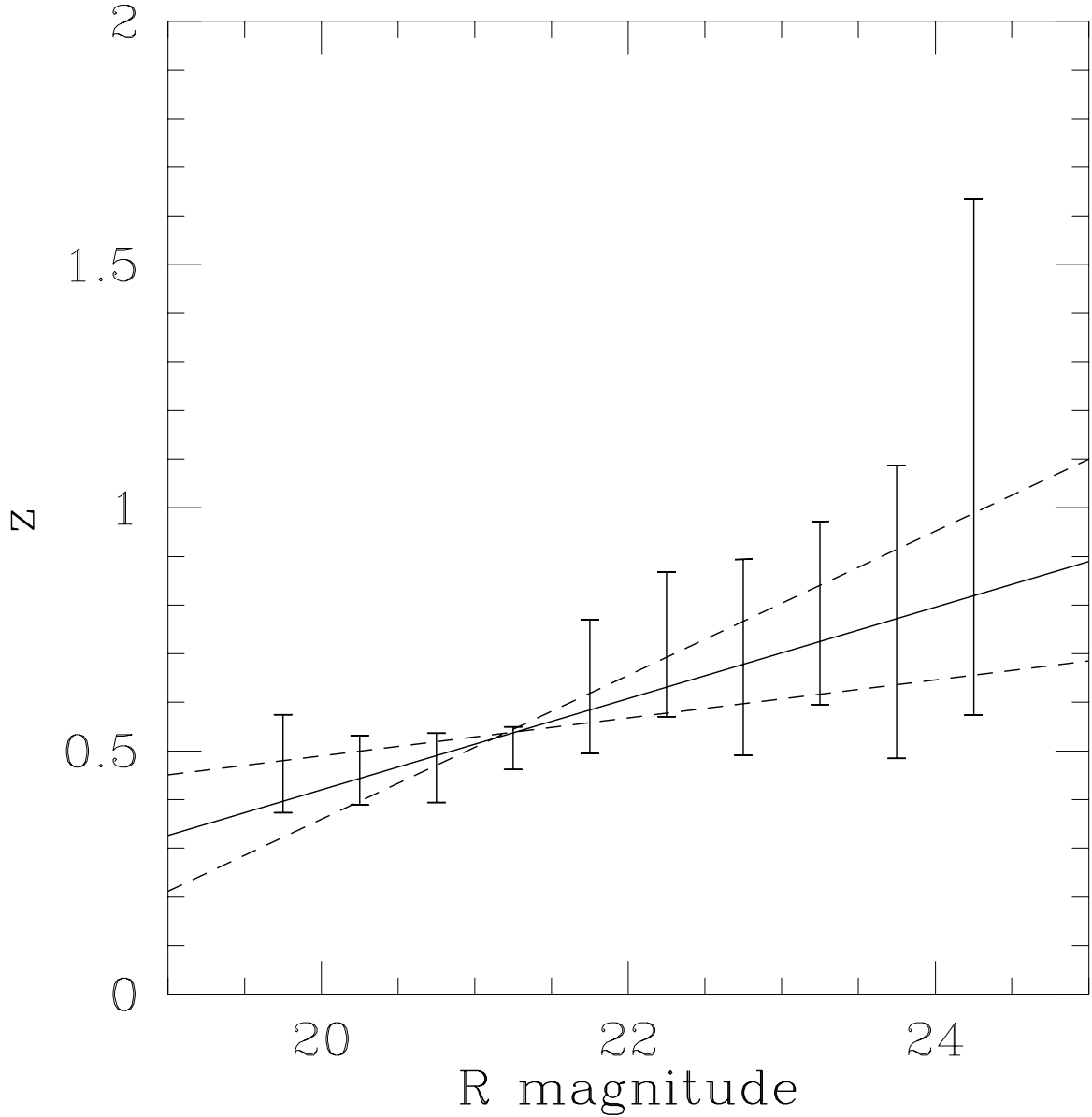} \caption{Redshift versus $R$ magnitude data from
the CalTech Faint Galaxy Redshift Survey. Objects whose redshift
fell one standard deviation from an initial bin average have been
excluded from this plot.  The error bars are $1\,\sigma$ error
bars for the recalculated bin average. The solid line corresponds
to a best fit redshift to magnitude relation that is used to
determine $\Sigma_{crit}$. The two dashed lines correspond to
$68\%$ confidence limits.\label{ms:fit:fig}}
\end{figure}

\begin{figure}
\plotone{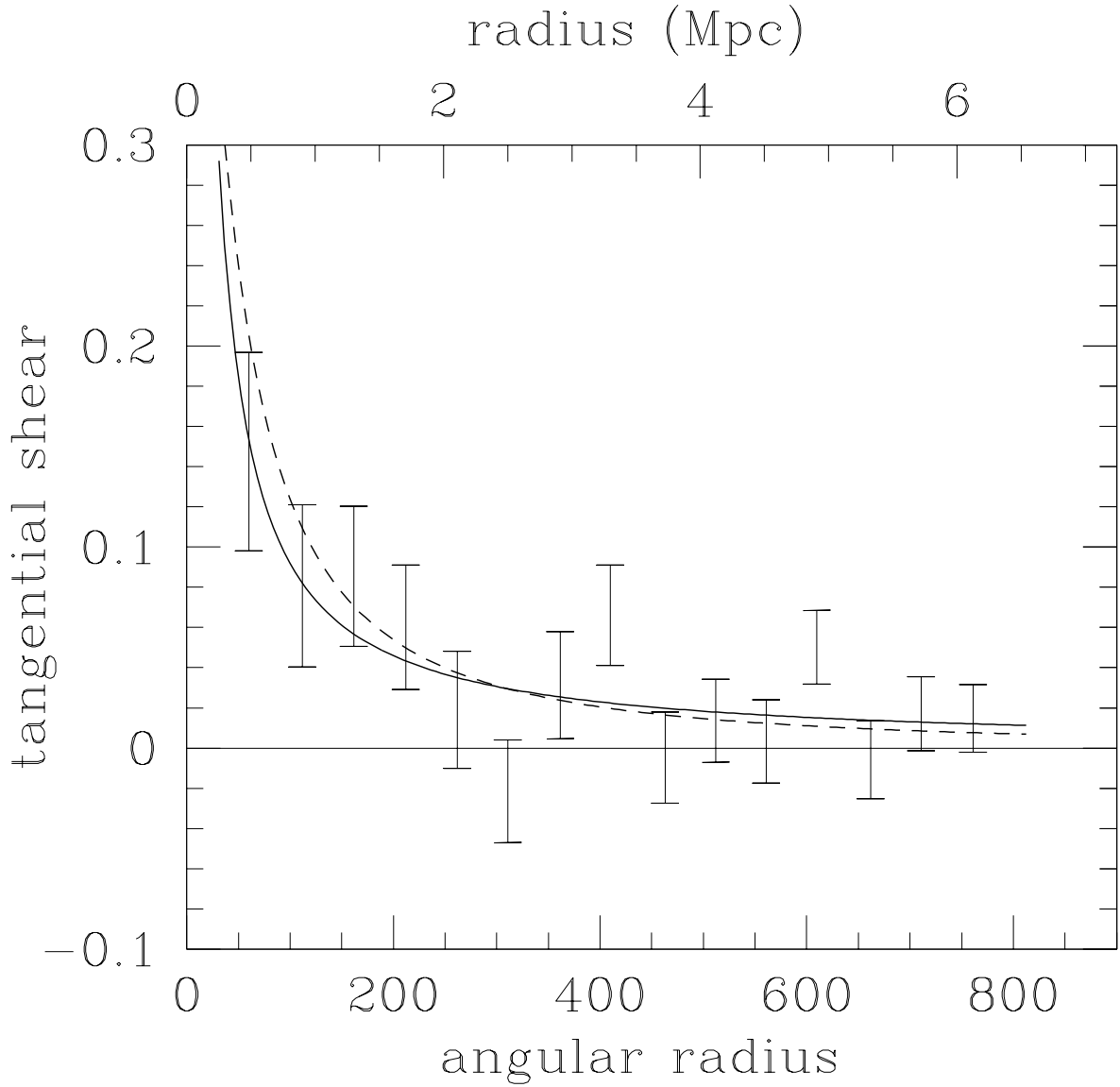} \caption{The tangential shear as a function of
angular radius with $1\sigma$ error bars is shown with results of
SIS (solid curve) and NFW (dashed curve) models using the best fit
parameters in an $\Omega_m = 0.3$, $\Omega_\Lambda = 0.7$
cosmological model.\label{fits}}
\end{figure}

\begin{figure}
\plottwo{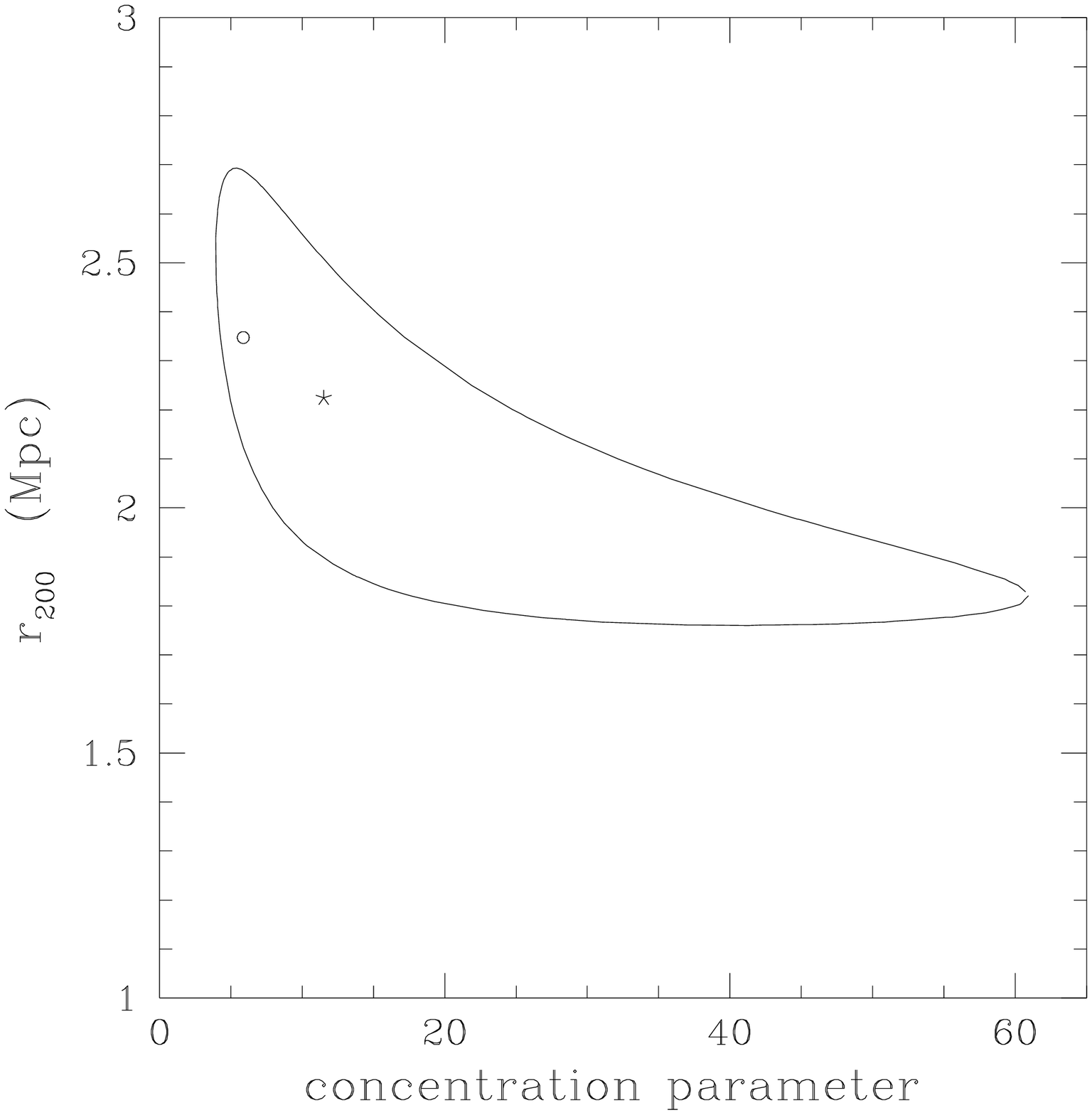}{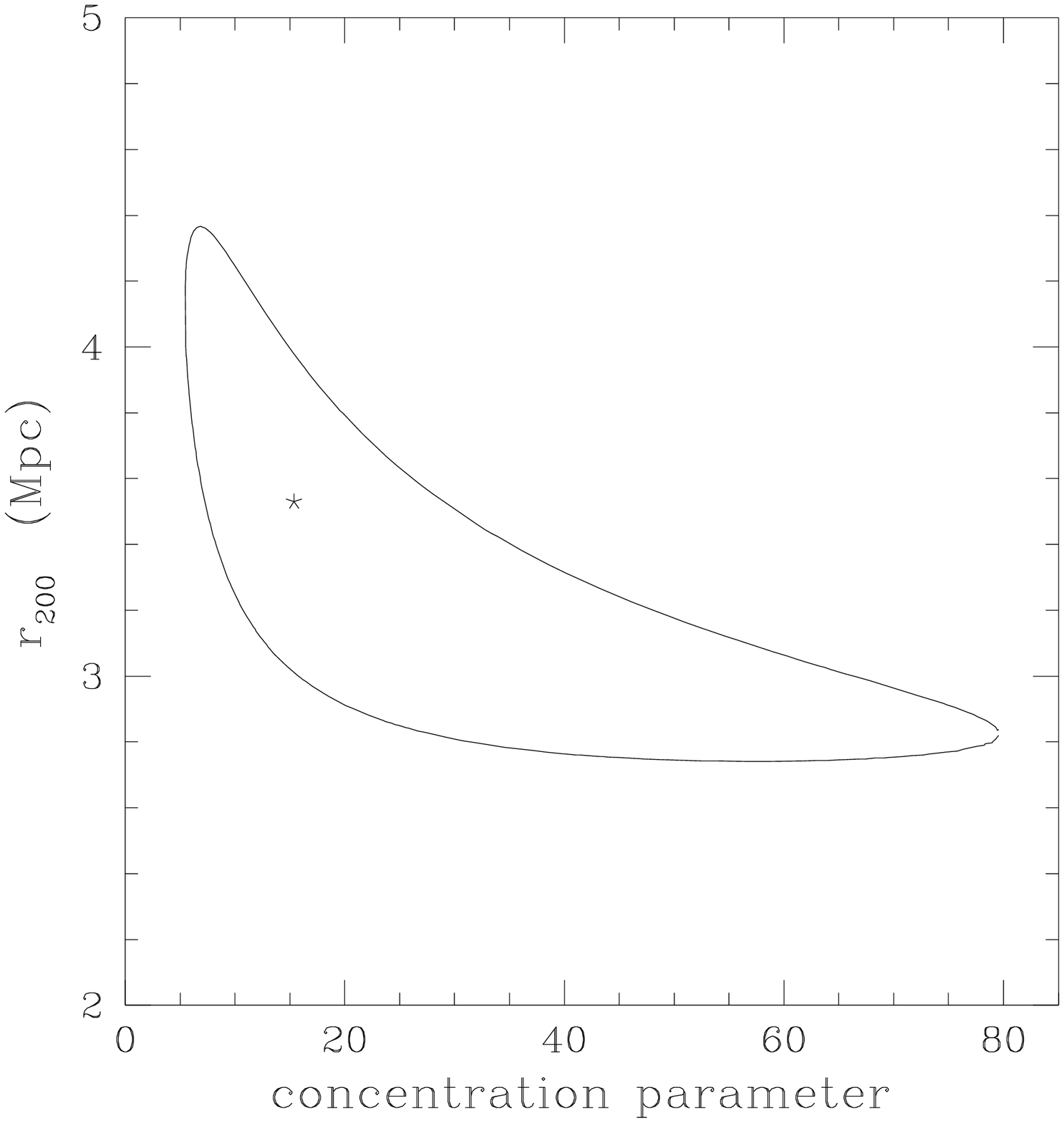} \caption{The $68\%$ confidence region
in the $c$-$r_{200}$ plane for the NFW model prediction to the
tangential shear of RXJ1347-1145 for $\Omega_m =1$,
$\Omega_\Lambda = 0$ (left) and $\Omega_m =0.3$, $\Omega_\Lambda =
0.7$ (right) cosmological models. The star corresponds to the
values that minimize $\chi^2$ for the current weak lensing data.
The circle represents the values obtained by Allen {\emph{et al}},
2002, from Chandra data. The projections of this error curve onto
the $c$ - $r_{200}$ axes leads to estimates of the error in the
integrated mass. \label{NFW_err} }
\end{figure}

\begin{figure}
\plotone{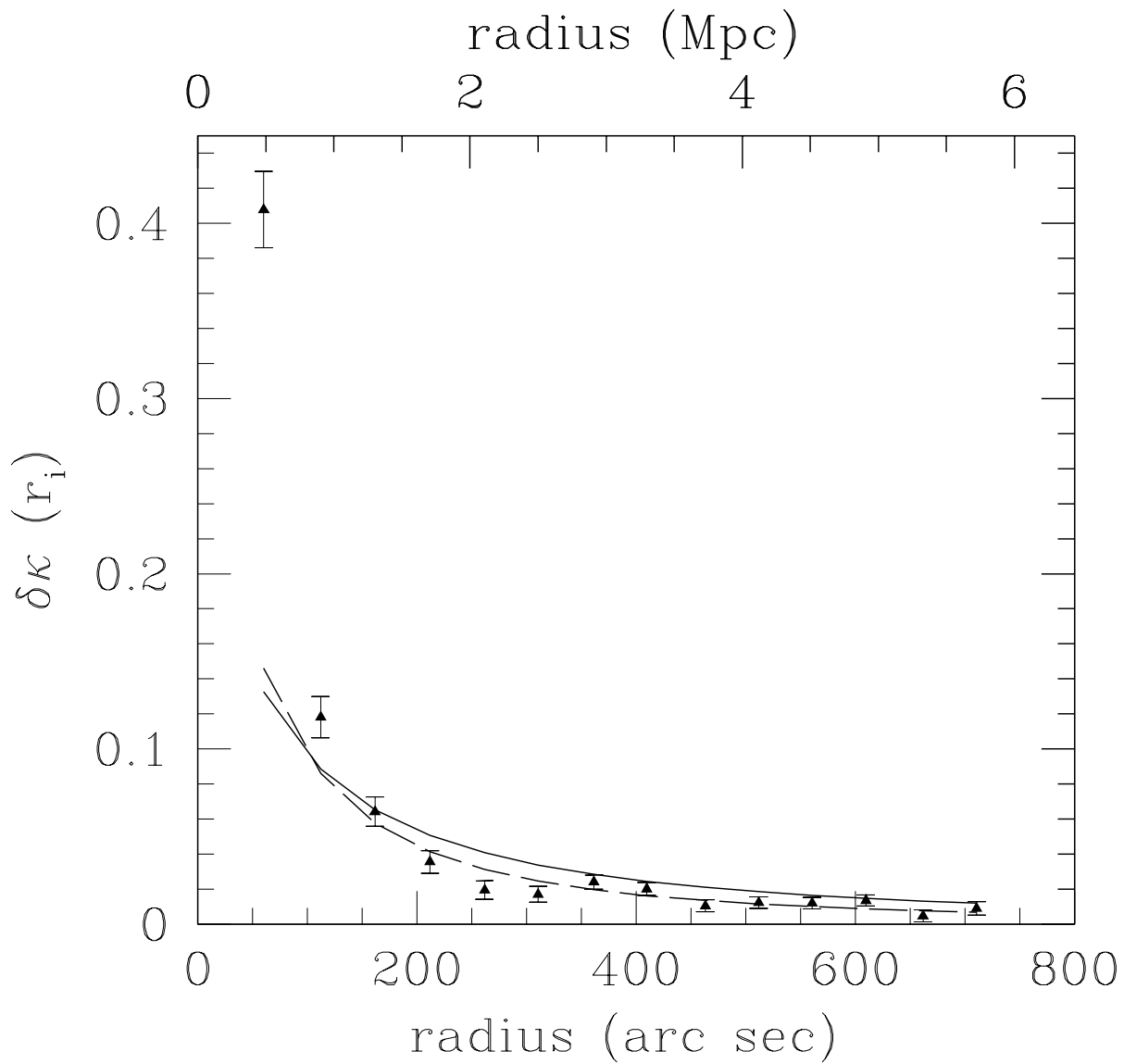} \caption{A densitometry measure ($\delta \kappa
= \bar\kappa (r<r_i) - \bar \kappa (r_i<r<r_o)$) of the mass of
RXJ1347--1145 with NFW (dashed line) and SIS (solid line) models
for the best fit parameters in the $\Omega_m = 0.3$,
$\Omega_\Lambda = 0.7$ cosmology.  Note that the error bars are
computed using only the number of galaxies in each bin, and are
thus non-independent underestimates of the error, especially in
the central bin. \label{mass:fig}}
\end{figure}

\begin{figure}
\plotone{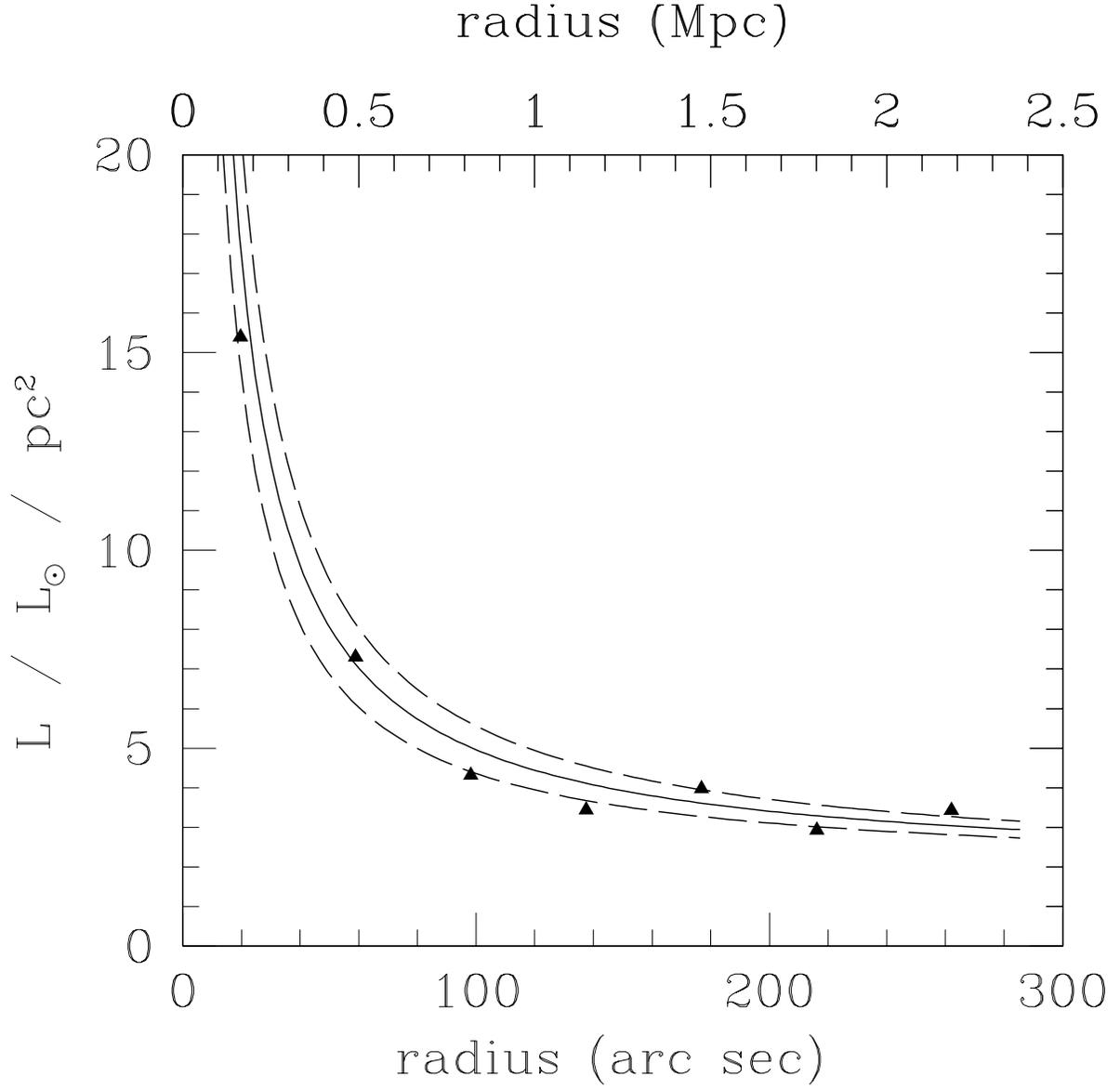} \caption{The luminosity per unit area in radial
annuli in units of solar luminosity per pc$^2$ (triangles) and a
best fit scaled mass density that is consistent with gravitational
shear (solid curve).  The dashed curves indicate $1\sigma$
variations in the scaled mass density.  The approximate ratio of
mass to light is $91$~M$_\odot / L_{R\odot}$.
 \label{lum:fig}}
\end{figure}

Future high resolution lensing studies of the interior region of
RXJ1347-1145 may yield interesting results.  We note that the
two-dimensional mass map presented in Fig.~\ref{fiatmap} does
suggest some structure, leading to the hope that high resolution
weak lensing studies of the central region may shed light on
possibility of a significant merging event.

\acknowledgments

This paper is based on observations made at the Cerro Tololo
Intra-American Observatory which is operated by the Association of
Universities for Research in Astronomy, under a cooperative
agreement with the National Science Foundation as part of the
National Optical Astronomy Observatories. This work partly
supported by NSF grant AST-0134743.


\clearpage
%
%

\begin{table}[hp] \begin{center}\begin{tabular}{llll}
\tableline Model & Cosmology & Parameters & Mass ($<1$ Mpc) \\
\tableline ~&~&~&~\\SIS & $\Omega_m = 1.0$, $\Omega_\Lambda = 0.0$
& $\sigma_v = 1500^{+140}_{-150}$ km s$^{-1}$ & $1.6\, \pm \, 0.3$
\\~&~&~&~\\SIS & $\Omega_m = 0.3$, $\Omega_\Lambda = 0.7$ & $\sigma_v =
1400^{+130}_{-140}$ km s$^{-1}$ & $1.4\, \pm \, 0.3 $
\\ ~&~&~&~\\ NFW & $\Omega_m = 1.0$, $\Omega_\Lambda = 0.0$ & $c = 11.5$ & ~\\
~&~& $r_s = 1.93$ Mpc & $1.5^{+0.6}_{-0.5}$
\\ ~&~&~&~\\NFW & $\Omega_m = 0.3$, $\Omega_\Lambda = 0.7$ & $c = 15.4$ &
~\\ ~&~& $r_s = 2.29$ Mpc & $1.4^{+0.5}_{-0.4}$
\\ \tableline
\end{tabular} \caption{SIS and NFW model best fit parameters and
total integrated masses within $1.0$ Mpc (in units of $10^{15}
M_\odot$) for two different cosmological models with $h = 0.5$ for
gravitational shear detected from objects with $20<R<26$. The
$\Omega_m = 1$ models are consistent with a previous weak lensing
study and the most recent Chandra X-ray observations. The
uncertainties in mass assume $\Sigma_{crit} = 8200 \,h\, M_\odot$
pc$^{-2}$ for the $\Omega_m = 1$ cosmology and $\Sigma_{crit} =
6000 \, h\, M_\odot$ for the $\Omega_m = 0.3$, $\Omega_\Lambda =
0.7$ cosmology. \label{model:table} }
\end{center}
\end{table}

\begin{table}[hp] \begin{center}\begin{tabular}{llll}
\tableline Model & Cosmology & Parameters & Mass ($<1$ Mpc) \\
\tableline ~&~&~&~\\SIS & $\Omega_m = 1.0$, $\Omega_\Lambda = 0.0$
& $\sigma_v = 1580^{+180}_{-210}$ km s$^{-1}$ & $1.8\, \pm \, 0.4$
\\~&~&~&~\\SIS & $\Omega_m = 0.3$, $\Omega_\Lambda = 0.7$ & $\sigma_v =
1550^{+180}_{-210}$ km s$^{-1}$ & $1.8\, \pm \, 0.4 $
\\ ~&~&~&~\\ NFW & $\Omega_m = 1.0$, $\Omega_\Lambda = 0.0$ & $c = 4.8$ & ~\\
~&~& $r_s = 0.584$ Mpc & $2.2^{+0.5}_{-0.5}$
\\ ~&~&~&~\\NFW & $\Omega_m = 0.3$, $\Omega_\Lambda = 0.7$ & $c = 6.8$ &
~\\ ~&~& $r_s = 0.691$ Mpc & $2.1^{+0.2}_{-0.3}$
\\ \tableline
\end{tabular} \caption{SIS and NFW model best fit parameters and
total integrated masses within $1.0$ Mpc (in units of $10^{15}
M_\odot$) for two different cosmological models with $h = 0.5$ for
gravitational shear detected from objects with $22<R<26$. The
uncertainties in mass assume $\Sigma_{crit} = 8000 \,h\, M_\odot$
pc$^{-2}$ for the $\Omega_m = 1$ cosmology and $\Sigma_{crit} =
6500 \, h\, M_\odot$ for the $\Omega_m = 0.3$, $\Omega_\Lambda =
0.7$ cosmology. \label{model2:table} }
\end{center}
\end{table}

\clearpage
%
%

\begin{thebibliography}{}

\bibitem[Allen et al.(2002)]{allen2002} Allen, S.W., Schmidt,
R.W., \& Fabian, A.C., 2002, \mnras, 335, 256A

\bibitem[Allen \& Fabian(1998)]{allen1998} Allen, S.W. \&
Fabian, A.C., 1998, \mnras, 297, L57

\bibitem[Bernstein \& Jarvis(2002)] {bernstein} Bernstein, G.M.
\& Jarvis, M., 2002, \aj, 123, 583B

\bibitem[Bertin \& Arnouts(1996)]{bertin} Bertin, E. \&
Arnouts, S., 1996, A\&A Sup Ser 117, 393

\bibitem[Cohen et al.(2000)]{FGRS} Cohen, J.G. et al., 2000, \apj, 538, 1

\bibitem[Cohen \& Kneib(2002)]{cohen-kneib} Cohen, J.G.
\& Kneib, J.P., 2002, \apj, 513, 524C

\bibitem[Ettori et al.(2001)]{ettori} Ettori. S.,
Allen, S.W., \& Fabian, A.C., 2001 \mnras, 322, 187

\bibitem[Fahlman et al.(1994)]{fahlman} Falman, G.G. et al., 1994
\apj, 436, 56

\bibitem[Fischer \& Tyson(1997)]{FT} Fischer, P., Tyson, J.A., 1997, \aj, 114 14

\bibitem[Fukugita et al.(1995)]{fuku} Fukugita et al., 1995 \pasp, 107,
945

\bibitem[Komatsu et al.(2001)]{komatsu} Komatsu, E. et al., 2001, \pasj, 53, 57

\bibitem[Khiabanian (2003)]{hossein} Khiabanian, H.
{\emph{personal communication}}

\bibitem[Miralda-Escude(1991)]{miralda} Miralda-Escude, J., 1991, \apj, 370, 1

\bibitem[Navarro et al.(1997)]{nfw} Navarro, J. F., Frenk, C. S.
\& White, S. D.M., 1997, \apj 490, 493

\bibitem[Pointecouteau et al.(2001)]{pointecouteau} Pointecouteau, E.,
Giard, M., Benoi, A., Desert, F.X., Bernard, J.P., Coron, N. \&
Lamarre, J.M., 2001, \apj, 552, 42

\bibitem[Press et al.(1995)] {nrc} Press, W.H. et al., 1995, {\emph{Numerical Recipes}},
Cambridge University Press

\bibitem[Schindler et al.(1995)]{schindler95} Schindler, S. et al., 1995, A\&A, 299L, 9

\bibitem[Schindler et al.(1997)]{schindler97} Schindler, S.,
Hattori, M., Neumann, D.M. \& B\"ohringer, H., 1997 A\&A, 317, 646

\bibitem[Schneider et al.(1992)]{ehlers} Schneider, P.,
Ehlers, J. \& Falco, E.E., 1992, {\emph{Gravitational Lensing,}}
Springer-Verlag

\bibitem[Shectman et al.(1996)]{camp} Shectman, S.A. et al., 1996
\apj, 470, 172s

\bibitem[Wittman et al.(2002)]{wittman} Wittman, D.E., et al.,
2002, SPIE, 4836, p. 21

\bibitem[Wright \& Brainerd(2000)]{wright} Wright, C.O.
\& Brainerd, T.G., 2000, \apj, 534, 34W



\end{thebibliography}
\end{document}